\documentclass[11pt]{article}
\usepackage[top=2cm,left=2cm,right=2cm]{geometry}
\usepackage[numbers,square]{natbib}

\usepackage[english]{babel}
\usepackage[latin1]{inputenc}
\usepackage{hyperref}
\usepackage{amssymb}
\usepackage{amsfonts}
\usepackage{amsmath}
\usepackage{fancyhdr}
\usepackage{epsfig}
\usepackage{graphics}
\usepackage{subfigure}
\usepackage{rotating}
\usepackage{lineno}
\usepackage{pstricks}
\usepackage{color}

            \let\d = \delta        
                  \let\m = \mu
\let\n = \nu                            \let\t = \tau

\newcommand{\muu}{\mu_{1}}
\newcommand{\nuu}{\nu_{1}}
\newcommand{\ku}{k_{1}}
\newcommand{\xu}{x_{1}}

\newcommand{\mud}{\mu_{2}}
\newcommand{\nud}{\nu_{2}}
\newcommand{\kd}{k_{2}}
\newcommand{\xd}{x_{2}}

\newcommand{\mut}{\mu_{3}}
\newcommand{\nut}{\nu_{3}}
\newcommand{\kt}{k_{3}}
\newcommand{\xt}{x_{3}}

\newcommand{\kq}{k_{4}}

\newcommand{\kc}{k_{5}}

\newcommand{\ks}{k_{6}}

\newcommand{\beq}{\begin{equation}}
\newcommand{\eeq}{\end{equation}}
\newcommand{\bea}{\begin{eqnarray}}
\newcommand{\eea}{\end{eqnarray}}

\newcommand{\nn}{\nonumber}
\newcommand{\spu}{\,\,\,}

\newcommand{\pd}{\partial}

\begin{document}
\begin{center}
\vspace{4.cm}
{\bf \large The Dilaton Wess-Zumino Action in 6 Dimensions from Weyl Gauging:\\ Local Anomalies and Trace Relations \\}
\vspace{1cm}
{\bf Claudio Corian\`{o}, Luigi Delle Rose, Carlo Marzo and Mirko Serino}

\vspace{1cm}

{Dipartimento di Matematica e Fisica "{\em Ennio De Giorgi}"\\ 
Universit\`a del Salento \\ and \\ INFN Lecce, Via Arnesano 73100 Lecce, Italy\\}
\vspace{0.5cm}

\begin{abstract}

We extend a previous analysis on the derivation of the dilaton Wess-Zumino (WZ) action in $d=4$, based on the method of Weyl gauging, 
to $6$ dimensions. As in the previous case, we discuss the structure of the same action in dimensional regularization using 
6-dimensional Weyl invariants, extracting the dilaton interactions in the most general scheme, with the inclusion of the local 
anomaly terms.    
As an application, we present the WZ action for the (2,0) tensor multiplet, which has been investigated in the past in the context of 
the $AdS_7/CFT_6$ holographic anomaly matching.
We then extend to $d=6$ the investigation of fully traced correlation functions of EMT's, formerly presented in 
$d=4$, showing that their hierarchy is functionally related only to the first 6 correlators. 
We give the explicit expressions of these in the most general scheme, up to rank-4.

\end{abstract}

\end{center}

\newpage

\section{Introduction} 

Anomaly-induced actions play a considerable role among effective field theories.
Simple instances of these types of actions are theories with chiral fermions in the presence of anomalous abelian symmetries 
\cite{Treiman:1986ep, Wess:1971yu, Faddeev:1984jp, Babelon:1986sv}, other examples involve conformal 
\cite{Duff:1977ay,Deser:1999zv, Deser:1993yx,Polyakov:1981re} and superconformal anomalies  \cite{Chaichian:2000wr}.

Direct computations of these actions can be performed in ordinary perturbation theory by the usual Feynman expansion at 1 loop, but alternative 
approaches are also possible. In fact, an action which reproduces the same 
anomaly at low energy can be constructed quite directly, just as a variational solution of the anomaly condition, without any reference to the 
diagrammatic expansion.
In gravity, typical examples are anomaly actions such as the Riegert action \cite{Riegert:1987kt}, or the 
Wess-Zumino (WZ) dilaton action \cite{Antoniadis:1992xu}, which reproduce the anomaly either with a non-local (Riegert) or with a local 
(WZ) effective operator, using a dilaton field in the latter case \cite{Codello:2012sn}. These types of actions are not unique, 
since possible contributions which are conformally invariant are not identified by the variational procedure. 
It should also be mentioned that a prolonged interest in these actions has been and is linked to the study of the irreversibility of the Renormalization 
Group (RG) flow in various dimensions (see for instance \cite{Jack:1990eb,Cappelli:1990yc, Anselmi:1997ys, 
Komargodski:2011vj,Yonekura:2012kb} and of the trace anomaly matching \cite{Schwimmer:2010za}, since Zamolodchikov's proof of his $c$-theorem in $d=2$ \cite{Zamolodchikov:1986gt}.
 
 A salient feature of some of these anomaly actions, particularly if formulated in a local form, as in the WZ case, is the inclusion 
of extra degrees of freedom compared to the original tree-level action. In the case of the chiral anomaly this additional degree of 
freedom is the axion $(\theta(x))$, which is linearly coupled to the anomaly functional in the form of a $(\theta/M) F\tilde{F}$ term 
- the anomaly coupling - with $F$ and $\tilde{F}$ denoting the field strength of the gauge field and its dual respectively. 
The anomaly interaction is accompanied by a new scale ($M$). This is the scale at which the anomalous symmetry starts to play a role 
in the effective theory. A large value of $M$, for instance, is then associated with a decoupling of the anomaly in the low energy theory. 
In the 1-particle irreducible (1PI) effective action this is obtained - in the chiral case - by allowing the mass of the fermions ($\sim M$)
that run in the anomaly loops to grow large. The underlying idea of keeping the anomaly interaction in the form of a local operator at low energy - 
such as the $(\theta/M) F\tilde{F}$ term - while removing part of the physical spectrum, is important in the study of 
the renormalization group (RG) flows of large classes of theories, both for chiral and for conformal anomalies. 

In the case of conformal anomalies \cite{Duff:1977ay}, which is the case of interest in this work, the pattern is similar to the 
chiral case, with the introduction of a dilaton field in place of the axion in order to identify the structure of the corresponding WZ action, and the inclusion of a conformal scale ($\Lambda$).  As in the chiral case, one of the significant 
features of the WZ conformal anomaly action is the presence of a linear coupling of the Goldstone mode of the broken symmetry (the 
dilaton) to the anomaly functional, but with a significant variant. In this case, in fact, this linear term has to be corrected by 
additional contributions, due to the non invariance of the anomaly functional under a conformal transformation. 

This procedure, which 
allows to identify the structure of WZ action, goes under the name of the {\em Noether method} (see for instance 
\cite{Coriano:2013xua}, \cite{Elvang:2012st}) and has to be iterated several times, due to the structure of the anomaly functional, 
before reaching an end. Given the fact that anomaly functional takes a different form in each space-time dimension, the anomaly action 
will involve interactions of the dilaton field of different orders in each dimension.

\subsection {Weyl gauging and the anomaly action}

In a previous paper \cite{Coriano:2013xua} we have investigated an alternative approach, useful for the computation of this action, 
which exploits the structure of the counterterms in dimensional regularization and their {\em Weyl gauging}, bypassing altogether the 
Noether procedure. This approach has been discussed in $d=4$ by several authors
\cite{Antoniadis:1992xu,Tomboulis:1988gw}, and in a cohomological context in \cite{Mazur:2001aa}.
 
In particular, we have shown how the complete hierarchy of correlators involving traces $(T)$ of the EMT's of any 
conformal field theory (CFT) in 4 dimensions is functionally related only to the first four correlators, 
with ranks from 1 to 4 ($ T, T^2,T^3,T^4 $), of the same theory. These are completely determined by the anomaly. 
It is the order of the dilaton interaction which determines the maximum rank of independent traced correlators necessary to fix the entire 
hierarchy.  In $d=6$, as we are going to show, the traces of the first 6 correlators ($T, T^2 ,\ldots, T^6$) are sufficient for this goal.

The extension of this construction to higher dimensions is interesting for several reasons. The WZ action in $d=6$ plays an important 
role in the study of the irreversibility of the RG flow of CFT's  from the ultraviolet (UV) to the infrared (IR) 
\cite{Komargodski:2011vj,Elvang:2012st} also for this specific dimensions. At the same time it plays an equally important role in the study of the AdS/CFT correspondence. An example is the investigation of the 
anomaly matching between conformal tensor multiplets on the six dimensional boundary and a stack of M5 branes of $AdS_7$ supergravity in the bulk 
\cite{Bastianelli:1999ab, Bastianelli:2000hi}. We will present, as an application of our formalism, the expression of the WZ action for 
this specific CFT realization in $d=6$. 
 
Our work, in the study that we present, follows rather closely the layout of our previous derivation of the WZ action by Weyl gauging in $d=4$, that we extend to 6 dimensions.
As in the 4-dimensional case, we derive the dilaton effective action by taking into account all the possible counterterms in the 
construction, which are identified in dimensional regularization within a general subtraction scheme. 
We also mention that  general results on the structure of the action in any even dimensions have been presented in 
\cite{Baume:2013ika}, using the general form of the Euler density and its conformal variation, which is sufficient to identify the 
nonlocal structure of the anomaly in a specific scheme, as we will specify below. However, the identification of the  
local contributions to the anomaly requires a separate effort, that we undertake in this work for $d=6$. 
 
This more general approach allows us to set a distinction between the nonlocal and local 
contributions to the anomaly and hence to the effective action, as in our former analysis of the $d=4$ case. 
We will start our investigation by reviewing the method of Weyl gauging, together with a brief discussion of the structure of 
correlation functions of traces of the energy-momentum tensor (EMT) for a generic CFT. In the past, the gauging has been discussed 
in various ways both in the context of extensions of the Standard Model \cite{Buchmuller:1988cj, Buchmuller:1987uc} 
and in cosmology, where it has been shown that the introduction of an extra scalar brings to a dynamical adjustment of the cosmological constant \cite{Tomboulis:1988gw}. Recent discussions of the role of the dilaton 
in quantum gravity can be found in \cite{Henz:2013oxa, Percacci:2011uf}. 

The extension of the gauging procedure from 4 
to 6 dimensions, in a general scheme, is quite demanding from the technical side, and is addressed starting from section 2, where we fix our 
conventions for the structure of the anomaly functional. This is expressed in terms of the generic coefficients $(c_1,c_2,c_3)$ and $a$, 
which describe the anomaly of any CFT in $d=6$. We thus identify the operators in the effective action that are responsible for the local 
anomaly, which are the analogue of the $R^2$ curvature term in $d=4$, explicitly establishing their connection to the part proportional to total 
derivatives. We then move to the analysis of the structure of the traced correlators and of their hierarchy, showing how to solve it in terms of the 
first  6 correlation functions. We have left to Appendix \ref{Geometrical} a discussion of some of the more technical steps.
Appendix \ref{2D} includes the consistency checks of the recursion relations satisfied by the traced correlators in $d=2$ and $d=4$, 
presenting the expressions of the first traced Green functions up to rank-6 in the two cases.  

\section{Definitions and conventions}

In this section we establish our conventions, which will be used throughout our computations, before coming to a description 
of the anomaly-induced action in $d=6$.  We define the generating functional of the theory $\mathcal{W}$  as
\beq
\mathcal{W}[g] = \int \mathcal D \Phi\, e^{- \mathcal S}\, ,
\eeq
where $\mathcal{S}$ is the generic euclidean action depending on the set of all the quantum fields ($\Phi$) and on the 
background metric ($g$). The EMT is given by
\beq \label{EMTvev}
\left\langle T^{\mu\nu}(x) \right\rangle 
= \frac{2}{\sqrt{g_x}}\, \frac{\delta \mathcal{W}[g]}{\delta g_{\mu\nu}(x)}
= \frac{2}{\sqrt{g_x}}\, \frac{\delta}{\delta g_{\mu\nu}(x)} \int\, \mathcal{D}\Phi\,  e^{-S}\, 
\eeq
which is symmetric and traceless for a conformal invariant theory. In (\ref{EMTvev}) $g_x\equiv \left|g_{\mu\nu}(x)\right|$ is the 
determinant of the metric tensor. 
For ordinary field theories, investigated in ordinary Minkowski 
space, this approach allows to identity a symmetric expression of the EMT, which is traceless in the presence of a scale invariant 
Lagrangian, modulo standard improvement terms for scalar fields that we will discuss next.
In the case of the Standard Model, this approach has been used to fix  the entire structure of the EMT 
in the $R_\xi$ gauge \cite{Coriano:2011zk}.

The equation of the conformal anomaly is expressed in terms of a functional $\mathcal{A}[g]$ 
which depends on the metric background $g$
\bea \label{TraceAnomaly}
g_{\mu\nu} \langle T^{\mu\nu} \rangle_g = \mathcal{A}[g]\, ,
\eea
and holds in any even dimensions.

The general structure of the trace anomaly equation for general even dimension $d$ is given by \cite{Deser:1993yx}
\beq
\label{anom}
\mathcal{A}[g] = 
\sum_{i} c_i\, \left( I_i + \nabla_\mu J^\mu_i \right) - (-1)^{d/2}\, a\, E_d \, ,
\eeq
where $\sqrt{g}\, I_i$ are conformal invariants, the analogous of the Weyl tensor squared in 4 dimensions, whose number increases 
with the dimension, whereas $E_d$
is the Euler density in $d$ dimensions.
The contribution coming from the Euler density is usually denoted as the $A$ part of the anomaly, while the rest is 
called the $B$ part.
The total derivative terms $\nabla^\mu J^i_\mu$ are known under the name of \emph{local anomaly contributions} and are sometimes omitted,
as they are scheme-dependent and absent if the $I_i$'s are expressed in generic $d$ dimensions.
They can be removed also by adding some local counterterms to the action, which are an intrinsic ambiguity of the dimensional
regularization scheme. \\ 
The specific expression of (\ref{anom}) for $d=6$ takes the form 
\beq \label{anom6D}
\mathcal{A}[g] = 
\sum_{i=1}^3 c_i\, \left( I_i + \nabla_\mu J^\mu_i \right) + a\, E_6\, , 
\eeq
where $\sqrt{g}\, I_i\, , (i=1,2,3)$, are the three conformal invariants available in 6 dimensions.
Our goal will be to determine the structure of the dilaton WZ action in the most general case, 
with the inclusions of the contributions related to these three conformal invariant terms.

\subsection{The conformal invariants and the Euler density in $d=6$ }\label{Basis}

To characterize the expansion of the scalars appearing in the trace anomaly equation,
we introduce the basis of scalars obtained from the Riemann tensor, its contractions and derivatives, which is given by 
\begin{table}[h]
$$
\begin{array}{lll}
 K_1    = R^3  &  
 K_2    = R\,R^{\mu\nu}\,R_{\mu\nu}  &  
 K_3    = R\,R^{\mu\nu\rho\sigma}\,R_{\mu\nu\rho\sigma}   \\
 K_4    = {R_\mu}^\nu\, {R_\nu}^\alpha\, {R_\alpha}^\mu   &  
 K_5    = R^{\mu\nu}\, R^{\rho\sigma}\, R_{\mu\rho\sigma\nu}  &  
 K_6    = R_{\mu\nu}\, R^{\mu\alpha\rho\sigma}\, {R^\nu}_{\alpha\rho\sigma}    \\
 K_7    = R_{\mu\nu\rho\sigma}\, R^{\mu\nu\alpha\beta}\, {R^{\rho\sigma}}_{\alpha\beta}  &
 K_8    = R_{\mu\nu\rho\sigma}\, R^{\mu\alpha\beta\sigma}\, {{R^\nu}_{\alpha\beta}}^\rho  &
 K_9    = R\square R    \\
 K_{10} = R_{\mu\nu}\square R^{\mu\nu} & 
 K_{11} = R_{\mu\nu\rho\sigma}\square R^{\mu\nu\rho\sigma} &  
 K_{12} = \pd_{\mu}R\, \pd^{\mu}R   \\
 K_{13} = \nabla_{\rho}R_{\mu\nu}\, \nabla^{\rho}R^{\mu\nu} & 
 K_{14} = \nabla_{\rho}R_{\mu\nu\alpha\beta}\, \nabla^{\rho}R^{\mu\nu\alpha\beta} &  
 K_{15} = \nabla_{\rho}R_{\mu\sigma}\, \nabla^{\sigma}R^{\mu\rho}   \\
\end{array}
$$
\end{table}
\\
in terms of which the Euler density takes the form 
\beq \label{Euler}
E_6 = K_1 - 12\, K_2 + 3\, K_3 + 16\, K_4 - 24\, K_5 - 24\, K_6 + 4\, K_7 + 8\, K_8  \, .
\eeq

Defining a Weyl transformation of the metric in the form 
\beq
g_{\mu\nu}(x) \to  e^{2\, \sigma(x)}\, g_{\mu\nu}(x),
\label{WeylT}
\eeq
the three Weyl invariants (modulo a $\sqrt{g}$ factor) in $d=6$ are given by the expressions 
(see Appendix \ref{Geometrical2} for their definitions in terms of the Weyl and Riemann tensors)
\bea \label{WeylInv6}
I_1
&=&
\frac{19}{800}\, K_1 - \frac{57}{160}\, K_2 + \frac{3}{40}\, K_3 + \frac{7}{16}\, K_4 
- \frac{9}{8}\, K_5 - \frac{3}{4}\, K_6 + K_8
\, ,\nn \\
I_2
&=&
\frac{9}{200}\, K_1 - \frac{27}{40}\, K_2 + \frac{3}{10}\, K_3 + \frac{5}{4}\, K_4 - \frac{3}{2}\, K_5 - 3\, K_6 + K_7
\, ,
\nn \\
I_3 &=&
\frac{1}{25}\, K_1 - \frac{2}{5}\, K_2 + \frac{2}{5}\, K_3 + \frac{1}{5}\, K_9 - 2\, K_{10} + 2\, K_{11} 
+ K_{13} + K_{14} - 2\,K_{15} \, .
\eea
It is easy to prove that for the three scalars defined above the products $\sqrt{g}\, I_i$ are Weyl invariant in $6$ dimensions, i.e.,
denoting with $\delta_W$ the operator implementing an infinitesimal Weyl transformation,
\beq
\delta_{W} I_i = -6 \, \sigma I_{i}\, .
\eeq

We also choose to define the Green function of $n$ EMT's in flat space in the completely symmetric fashion as 
\beq \label{NPF}
\langle T^{\mu_1\nu_1}(x_1)\ldots T^{\mu_n\nu_n}(x_n)\rangle 
\equiv
\frac{2^n}{\sqrt{g_{\xu}}\ldots \sqrt{g_{x_n}}}
\frac{\delta^n \mathcal{W}[g]}{\delta g_{\mu_1\nu_1}(\xu)\ldots \ldots\delta g_{\mu_n\nu_n}(x_n)}
\bigg|_{g_{\mu\nu}=\delta_{\mu\nu}}. \, 
\eeq
We denote the functional derivatives with respect to the metric of generic functionals, in the limit of a flat background, as
\bea \label{funcder}
\left[f(x)\right]^{\muu\nuu\dots\mu_{n}\nu_{n}}(\xu,\dots,x_n) 
\equiv
\frac{\delta^n\, f(x)}{\delta g_{\mu_n\nu_n}(x_{n}) \, \ldots\, \delta g_{\muu\nuu}(\xu)}
\bigg|_{g_{\mu\nu}=\delta_{\mu\nu}} 
\eea
and the corresponding expression with traced indices
\beq
\left[f(x)\right]^{\muu\dots\mu_n}_{\spu\muu\dots\mu_n}\left(\xu,\xd,\dots,x_n\right)
\equiv \delta_{\muu\nuu}\dots\delta_{\mu_{n}\nu_{n}}\,
\left[f(x)\right]^{\muu\nuu\dots\mu_{n}\nu_{n}}\left(\xu,\dots,x_n\right)\, ,
\eeq
where the curved euclidean metric $g_{\mu\nu}$ is replaced by $\delta_{\mu\nu}$. 

By functional differentiations of the anomaly equation (\ref{TraceAnomaly}), one generates an infinite hierarchy of 
equations satisfied by the correlation functions of multiple traces of the EMT in the form  
\bea
\label{hier}
\left\langle T(\ku) \, \dots \, T(k_{n+1})\right\rangle
&=&
2^n\, \left[\sqrt{g}\, \mathcal A \right]^{\muu\dots\mu_n}_{\spu\muu\dots\nu_n}\left(\ku,\dots,k_{n+1}\right)
\nn \\
&&
-\, 2 \sum_{i=1}^{n} \left\langle T(\ku)\dots T(k_{i-1})T(k_{i+1})\dots T(k_{n+1}+k_i) \right\rangle, \, 
\eea
which indicate the existence of an open hierarchy. As we have shown in our analysis of the $d=4$ case, this hierarchy can be 
completely identified just by a certain number of correlators, which in this case corresponds only to the first 6. However, as we 
have pointed out above, the number of traced correlators required to identify the hierarchy is related to the order of the dilaton 
interaction in the effective action.

In the expression above we have introduced the notation $T \equiv {T^\mu}_\mu$ to denote the trace of the EMT. All 
the momenta characterizing the vertex are taken as incoming, as specified in Appendix \ref{Geometrical}.

The identity (\ref{hier}) relates a $(n+1)$-point correlator to correlators of order $n$, together with the completely 
traced derivatives of the anomaly functionals $\sqrt{g}\,I_{i}, \sqrt{g}\,E_6$ and $\sqrt{g}\,\nabla_\mu J^\mu_i$. 
For $\sqrt{g}\,I_i$, which is a conformal invariant, they are identically zero.
For $\sqrt{g}\,E_6$ these are non vanishing at any arbitrary order $n \geq 4$, 
while $\sqrt{g}\,\nabla_\mu J^\mu_i$ contribute also to the trace of lower order functions.
In particular, as shown above, $\nabla_\mu J^\mu_{1/2}$ are at least quadratic in the Riemann tensor, so that they give
non-vanishing contributions from order $3$ onwards, whereas $\nabla_\mu J^\mu_3$ contains a term which is linear in $R$ 
and thus contributes a non-vanishing trace to the two-point function.

\section{Weyl gauging : overview}

\subsection{Weyl gauging for scale invariant theories}

The procedure of Weyl gauging defines a consistent framework useful to identify the coupling of a dilaton to the fields of a given
Lagrangian. It can be implemented starting from a Lagrangian defined on a flat metric background, but written in a diffeomorphic invariant way
(i.e. by using curvilinear coordinates), and introducing an appropriate new field which takes a role similar to an abelian gauge field. This allows to define a new Lagrangian which is
diffeomorphic and Weyl invariant in curved space. At a second stage this new degree of freedom can be made dynamical with the inclusion of a
kinetic term. As we are going to see, the transformation property of this new field, which can be traded for the gradient of a dilaton, together with the requirement of Weyl 
invariance, forces its kinetic term to a unique form. This is obtained performing a non linear field redefinition in the Lagrangian of a conformally coupled scalar.
The approach brings to the construction of a Weyl invariant Lagrangian which is conformally invariant in the flat limit. 
We are going to summarize these points below, illustrating explicitly the method in the simpler case of a scalar theory.

For a Lagrangian in flat space written in a diffeomorphic invariant form, scale invariance is equivalent to global Weyl invariance. 
The equivalence can be shown quite straightforwardly \cite{Iorio:1996ad} by rewriting a scale transformation acting on 
the coordinates of flat space and the fields $\Phi$
\bea \label{FlatScaling}
x^\mu &\to& {x'}^{\mu}=e^\sigma x^\mu \nn\\
\Phi(x)&\to& \Phi'(x')=e^{-d_\Phi \sigma}\Phi(x)
\eea
in terms of a rescaling of the Vielbein and of the  matter fields 
\bea \label{CurvedScaling}
{V}_{a\,\rho}(x) &\to&  e^{\sigma}\, V_{a\,\rho}(x), \nn\\
\Phi(x)&\to& e^{-d_\Phi \sigma}\Phi(x)
\eea
but leaving the coordinates of the field $\Phi$ invariant. We have denoted with $d_\Phi$ its scaling dimension.
Obviously, once we move to a curved metric background, it is natural to promote the global scaling parameter $w = e^\sigma$ 
to a local function, and modify the theory so that the transformation laws of the Vielbein and matter fields $(\Phi)$
\bea 
\label{WeylTransf}
{V'}_{a\,\rho}(x)  &=&  e^{\sigma(x)}\, V_{a\,\rho}(x)\, , \nn \\
\Phi'(x)           &=&  e^{-d_{\Phi}\, \sigma(x)}\, \Phi(x)\, 
\eea
leave the fundamental Lagrangian invariant.  
For a free scalar theory 
\beq
\frac{1}{2}\, \int d^d x \sqrt{g}\, g^{\mu\nu}\, \partial_{\mu}\phi\,  \partial_\nu \phi , 
\label{kin}
\eeq
the derivative terms are modified as for an abelian gauge field
\beq \label{DerTransf}
\pd_\mu \rightarrow \pd^W_\mu = \pd_\mu - d_{\phi}\, W_{\mu}\, ,
\eeq
where $W_{\mu}$ is a vector gauge field that transforms under Weyl scaling as
\beq \label{WeylGaugeField}
W_{\mu} \rightarrow W_{\mu} - \pd_\mu \sigma\, .
\eeq
In the case of a covariant derivative acting on a spin-1 field $v_\mu$, the Weyl and diffeomorphic covariant derivative 
is found by adding to (\ref{DerTransf}) the modified Christoffel connection
\beq \label{ModChristoffel}
\hat\Gamma^\lambda_{\mu\nu} =
\Gamma^\lambda_{\mu\nu} + {\delta_\mu}^\lambda\, W_\nu + {\delta_\nu}^\lambda\, W_\mu - g_{\mu\nu}\, W^\lambda\, ,
\eeq
which is Weyl invariant. The method follows closely the gauging of a typical abelian theory, by defining
\bea \label{WeylChristoffel}
\nabla^W_\mu v_\nu &=& \pd_\mu v_\nu - d_v\, W_\mu v_\nu - \hat\Gamma^\lambda_{\mu\nu} v_\lambda\,\nn\\
\nabla^W_\mu v_\nu  &\rightarrow& e^{- d_v\sigma(x)}\, \nabla^W_\mu v_\nu \, .
\eea
The extension to the fermion case is obtained by the relation
\beq \label{WeylSpinConnection}
\nabla_{\mu} \rightarrow \nabla^W_\mu = \nabla_{\mu} - d_\psi\, W_{\mu} + 2\, {\Sigma_\mu}^\nu\, W_\nu\, , \quad 
\Sigma^{\mu\nu} \equiv {V_a}^\mu\, {V_b}^\nu \Sigma_{ab}\, ,
\eeq
where we have denoted with $d_\psi$ the scaling dimension of the spinor field ($\psi$) and with 
$\Sigma_{ab}$ the spinor generators of the Lorentz group. 

If we Weyl-gauge the scalar action (\ref{kin}) according to the prescriptions 
in (\ref{DerTransf}) and (\ref{WeylGaugeField}) we obtain
\bea \label{WeylGaugeFieldScalar}
S_{\phi,W}
&=& 
\frac{1}{2}\, \int d^dx\, \sqrt{g}\,  g^{\mu\nu}\, \pd^W_\mu \phi\, \pd^W_\nu \phi
= \frac{1}{2}\, \int d^dx\, \sqrt{g}\,  g^{\mu\nu}\,\bigg( \pd_\mu - \frac{d-2}{2}\, W_\mu \bigg)\,\phi\,
\bigg( \pd_\nu - \frac{d-2}{2}\, W_\nu \bigg)\, \phi \nn \\
&=&
\frac{1}{2}\, \int d^dx\, \sqrt{g}\,  g^{\mu\nu} \bigg\{ \pd_\mu \phi\, \pd_\nu\phi
- \frac{d-2}{2}\, \bigg( \phi\,W_\mu\,\pd_\nu\phi + \phi\,W_\nu\,\pd_\mu \phi - \frac{d-2}{2}\, W_\mu\,W_\nu\,\phi^2  \bigg)
\bigg\}
\eea
which, using $\phi\, \pd_{\mu}\phi = 1/2\, \pd_\mu \phi^2$ and integrating by parts, can be written as
\beq
\mathcal S_{\phi,W} = 
\frac{1}{2}\, \int d^dx\, \sqrt{g}\,  g^{\mu\nu}\, \bigg(\pd_\mu \phi\, \pd_\nu\phi 
                                                          + \phi^2\, \frac{d-2}{2}\, \Omega_{\mu\nu}(W) \bigg)\, ,
\label{gauged}
\eeq
where we have introduced
\beq \label{Omega}
\Omega_{\mu\nu}(W) = \nabla_\mu W_\nu - W_\mu\,W_\nu + \frac{1}{2}\,g_{\mu\nu}\, W^2 \, .
\eeq
The result of this procedure is a Weyl invariant Lagrangian in which the Weyl variation of the 
ordinary kinetic term of $\phi$ is balanced by the variation of the $\Omega$ term.
One can also render $W_{\mu}$ dynamical by the inclusion of a kinetic term built out of an appropriate field strength 
\beq
F^W_{\mu\nu}\equiv\partial_{\mu} W_{\nu} - \partial_{\nu} W_{\mu}
\eeq
which is manifestly Weyl invariant.
 
A question that arises is whether it is possible to build a Weyl invariant theory without having to introduce an additional gauge field $W_\mu$ 
at all. As discussed in \cite{Iorio:1996ad}, this is possible if and only if, having performed the Weyl gauging, $W_\mu$ appears in the 
gauged action only in the combination given by $\Omega_{\mu\nu}(W)$. 
In fact, having observed that under a finite Weyl transformation ($\delta_W$)
the variation of $\Omega_{\mu\nu}(W)$ coincides, modulo a factor, with the variation of a particular combination of the Ricci tensor 
and the scalar curvature, i.e.
\bea
\delta_W \Omega_{\mu\nu}(W) 
&=& 
\frac{1}{2-d}\, \delta_W S_{\mu\nu}\, , 
\nn \\
S_{\mu\nu}
&=&
\frac{1}{2-d}\, \bigg( R_{\mu\nu} - \frac{1}{2\,(d-1)}\,g_{\mu\nu}\,R \bigg)\, ,
\eea
one obtains a new Weyl invariant action via the replacement
\beq
\Omega_{\mu\nu}(W) \rightarrow \frac{1}{2-d}\, S_{\mu\nu} \, .
\eeq
Doing so in the Weyl gauged action of the scalar field 
(\ref{gauged}), the latter takes the form
\beq \label{ScalarImproved}
\mathcal S_{\phi,\, imp} = \frac{1}{2}\, \int d^dx\, \sqrt{g}\, \bigg( g^{\mu\nu}\,\pd_\mu \phi\, \pd_\nu\phi 
- \frac{1}{4}\,\frac{d-2}{d-1}\, R\,\phi^2 \bigg)\,  ,
\eeq
which is the action of a conformally coupled scalar.
The procedure of rendering the theory Weyl invariant through such supplementary couplings to the Ricci tensor is called \emph{Ricci gauging} 
\cite{Iorio:1996ad}.

A second possibility is to maintain the expression of $W_\mu$, with new interactions induced by the Weyl gauging, but 
now identified with the gradient of a dilaton field, 
 \beq
 \label{dil}
 W_\mu(x) = \frac{\partial_{\mu} \tau(x)}{\Lambda}. 
 \eeq
This second choice offers an interesting physical interpretation - in the flat limit - in connection with the breaking
of the conformal symmetry, related to the conformal scale $\Lambda$, as we will shortly point out below.  Notice that in this second case the 
$\Omega(\partial_\mu \tau/\Lambda)$ term generates non trivial cubic and quartic interactions between
the original scalar and the dilaton 
\beq \label{Omegatau}
\Omega\left(\frac{\pd\tau}{\Lambda}\right) = 
\frac{\nabla_\mu \pd_\nu\tau}{\Lambda}  - \frac{\pd_\mu\tau\,\pd_\nu\tau}{\Lambda^2}
+ \frac{1}{2}\,g_{\mu\nu}\, \frac{\left(\pd\tau \right)^2}{\Lambda^2},
\eeq
which bring (\ref{gauged}) to the form
\beq
\mathcal S_{\phi,\pd\tau} = 
\frac{1}{2}\, \int d^dx\, \sqrt{g}\,  g^{\mu\nu}\, \left(
\pd_\mu \phi\, \pd_\nu\phi  + \frac{d-2}{2}\, \phi^2\, \frac{\Box\tau}{\Lambda} 
+ \left(\frac{d-2}{2}\right)^2 \, \phi^2\,  \frac{\left(\pd\tau\right)^2}{\Lambda^2} \right) \, .
\label{gaugedtau}
\eeq
As the field strength $F^W$, on account of (\ref{dil}) is obviously zero, the dilaton can be rendered dynamical only via a nonlinear realization of its kinetic term.
This is achieved by introducing  a conformally coupled scalar field $\chi$ and imposing the field redefiniton 
\beq
\chi(\tau) \equiv \Lambda^{\frac{d-2}{2}}\, e^{-\frac{(d-2)\,\tau}{2\Lambda}}.
 \label{chitau}
 \eeq
 At this point, the dynamics of the combined scalar/dilaton/graviton system is described by the Weyl invariant action
 \beq
 \mathcal{S}= S_{\chi(\tau), imp} + S_{\phi,\partial \tau} \, ,
 \eeq
having combined (\ref{ScalarImproved}), where $\phi$ is replaced by $\chi$, and (\ref{gaugedtau}). 
The  kinetic action for $\chi$, $S_{\chi(\tau), imp}$, takes the form
\beq \label{DilatonKinetic}
\mathcal S_{\chi(\tau), imp} = \frac{\Lambda^{d-2}}{2}\, \int d^dx\, \sqrt{g}\, e^{-\frac{(d-2)\,\tau}{\Lambda}}\, \bigg( 
\frac{(d-2)^2}{4\,\Lambda^2}\, g^{\mu\nu}\,\pd_\mu \tau\, \pd_\nu\tau - \frac{1}{4}\, \frac{d-2}{d-1}\, R \bigg) \, ,
\eeq
which, for the particular case $d=6$ of interest in this work, reduces to
\beq \label{KinTau}
\mathcal S_{\chi(\tau), imp} = 
\int d^6 x\, \sqrt{g}\, e^{-\frac{4\,\tau}{\Lambda}}\, 
\bigg( 2\, \Lambda^2\, g^{\mu\nu}\, \pd_\mu \tau\, \pd_\nu\tau - \frac{\Lambda^4}{10}\, R \bigg)\, .
\eeq
The Weyl gauging, as we have described it so far, is possible only when we take as a starting point a scale invariant Lagrangian, with dimensionless constants.
Things are different when an action is not scale invariant in flat space, and in that case the same gauging requires some extra steps. We illustrate this point below and discuss the modification of the procedure outlined above, by considering again a scalar theory as an example. This approach exemplifies a situation which is typical in theories with spontaneous breaking of the ordinary gauge symmetry, such as the Standard Model. 
 
 \subsection{Weyl gauging for non scale invariant theories } 
We consider a free scalar theory with a mass term 
\beq
\label{tre}
\mathcal{S}_2  = \frac{1}{2}\, \int d^d x \sqrt{g}\, \left(g^{\mu\nu}\,\partial_{\mu}\phi\, \partial_\nu \phi + m^2\, \phi^2 \right) \, .
\eeq
Scale invariance is lost, but it can be recovered. 
There are two ways to promote this action to a scale invariant one. The first is simply to render the mass term dynamical 
\beq
 \label{redef}
 m\to m\, \frac{\Sigma}{\Lambda}\, ,
 \eeq
using a second scalar field, $\Sigma$. The action (\ref{tre}), 
with the replacement (\ref{redef}), can be extended with the inclusion of the 
kinetic term for $\Sigma$.  The inclusion of $\Sigma$ and the addition of two conformal couplings (i.e. of two Ricci gaugings) both for 
$\phi$ and $\Sigma$ brings to the new action
\beq
\label{tre1}
\mathcal{S}^{\Sigma}_2  = \int d^d x \sqrt{g}\, \left[ \frac{1}{2}\,g^{\mu\nu}\, 
\bigg(\partial_{\mu}\phi\, \partial_\nu \phi + \partial_{\mu}\Sigma \, \partial_\nu \Sigma \bigg)
+ \frac{1}{2}\, m^2\, \frac{\Sigma^2}{\Lambda^2}\, \phi^2
+ \frac{1}{4}\, \frac{d-2}{d-1}\,R\, \bigg(\phi^2 + \Sigma^2\bigg) \right]\, ,
\eeq
which is Weyl invariant in curved space. 
These types of actions play a role in the context of Higgs-dilaton mixing in conformal invariant extension of the Standard 
Model, where $\phi$ is replaced by the Higgs doublet and $\Sigma$ is assumed to acquire a vacuum expectation value (vev)
which coincides with the conformal breaking scale $\Lambda$  ($\langle \Sigma \rangle=\Lambda$) 
(see for instance \cite{Coriano:2012nm}). The mixing is induced by a simple extension of (\ref{tre1}), where the mass term is 
generated via the scale invariant potential 
\beq
\mathcal{S}_{pot}= 
\lambda\, \int d^4 x\, \sqrt{g}\, \left( \phi^2 - \frac{\mu^2}{2\,\lambda}\frac{\Sigma^2}{\Lambda^2} \right)^2 \,  
\label{example}
\eeq
(with $m=\mu$). This choice provides a clear example of a Weyl invariant Lagrangian that allows a spontanous breaking of the $Z_2$ symmetry of the scalar sector $\phi$, following the breaking of the conformal symmetry ($\langle\Sigma \rangle=\Lambda, \textrm{with} \langle \tau\rangle=0$). The theory is obviously Weyl invariant  
(see the discussion in \cite{Coriano:2012nm}), but the contributions proportional to the Ricci scalar $R$ do not survive, obviously, in the flat limit.  

The approach to Weyl gauging of a non scale invariant Lagrangian briefly described above is not unique. In fact, a second alternative in the construction of a Weyl invariant Lagrangian in curved space, starting 
from (\ref{tre}), is to use the compensation procedure, 
which amounts to the replacements 
\bea \label{Compensate}
 m &\to& m \, e^{- \tau/\Lambda}\, ,\nn \\
 g_{\mu\nu}&\to& \hat{g}_{\mu\nu}\equiv g_{\mu\nu}\, e^{-2 {\tau/\Lambda }} \, \nn \\
 \phi&\to & \hat{\phi}\equiv \phi \, e^{\tau/\Lambda} \, , \nn\\
 \partial_\mu \phi &\to& \partial_\mu \hat{\phi}= e^{\tau/\Lambda}\, \partial^W_\mu\, \phi  \, ,
\qquad \textrm{with} \qquad W_\mu = \frac{\partial_{\mu}\tau}{\Lambda} \, ,
 \eea
giving an action of the form
\beq
\hat{\mathcal{S}}_2 
\equiv
\mathcal{S}_2(\hat{g},\hat{\phi})= 
\frac{1}{2}\, \int d^4 x\,  \sqrt{g}\,  \bigg[ g^{\mu\nu}\,\partial_{\mu}\phi\, \partial_\nu \phi
+ g^{\mu\nu}\,\Omega_{\mu\nu}\left(\frac{\partial \tau}{\Lambda}\right)\phi^2 + m^2\, e^{-2\,\tau/\Lambda}\, \phi^2 
\bigg] \, ,
\label{newflat}
\eeq
where $\Omega(\pd\tau/\Lambda)$ was defined in (\ref{Omegatau}). 
Also in this case, the compensator $\tau$ becomes a dynamical dilaton field by adding to 
$\hat{\mathcal{S}}_2$ the kinetic contribution of a conformally coupled scalar (\ref{DilatonKinetic}), 
obtaining the total action
\beq 
\mathcal{S}_T \equiv \hat{\mathcal{S}}_2 + \mathcal{S}_{\chi(\tau), imp}\, . 
\eeq
Notice that in this case we choose not to require the Ricci gauging of the $\Omega\left(\pd\tau/\Lambda \right)$ term in $\hat{\mathcal{S}}_2$, but we leave it as it is, thereby generating additional interactions between the dilaton and the scalar $\phi$ in flat space. Obviously, also following this second route, we can incorporate spontaneous breaking of the $Z_2$ symmetry of the 
$\phi$ field after the breaking of conformal invariance (with $\langle\Sigma\rangle=\Lambda)$. This is obtained, as before, by the inclusion of the potential (\ref{example}).

In this second approach the $\Omega(\partial \tau/\Lambda)$ terms are essential in order to differentiate between the two residual dilaton interactions in flat space.
In the context of Weyl invariant extensions of the Standard Model, such terms are naturally present in the analysis  
of \cite{Buchmuller:1988cj}.
\section{ Weyl gauging of the renormalized action}

The WZ anomaly action, as we have already mentioned above, is derived from the Weyl gauging of the renormalized action, defined 
as
\beq
\hat\Gamma_{\textrm{ren}}[g,\tau] \equiv \Gamma_{0}[g,\tau] + \Gamma_{\textrm{Ct}}[\hat{g}]\, ,
\eeq
in terms of a Weyl invariant contribution $\Gamma_{0}[g,\tau] $ and of a local counterterm $\Gamma_{\textrm{Ct}}[\hat{g}]$.
The Wess-Zumino action is then identified from the relation
\beq
\label{coc}
\hat\Gamma_{\textrm{ren}}[g,\tau] = \Gamma_{\textrm{ren}}[g,\tau] - \Gamma_{\textrm{WZ}}[g,\tau].
\eeq
Here $\Gamma_{WZ}[g,\tau]$ is the Wess-Zumino action, whose Weyl variation equals the trace anomaly. 
Notice that ${\hat{\Gamma}}_{\textrm{ren}}[{g,\tau}]$, as one can immediately 
realize, is Weyl invariant by construction, being a functional only of $\hat{g}$.
The Weyl invariant terms may take the form of any scalar contraction of $\hat R_{\mu\nu\rho\sigma}$, $\hat R_{\mu\nu}$ and $\hat R$
and can be classified by their mass dimension, such as
\beq
\mathcal{J}_n[\hat{g}] \sim \frac{1}{\Lambda^{2\,n - d}}\int d^d x \sqrt{\hat{g}}\hat{R}^n \, ,
\eeq
and so forth. In principle, all these terms can be included into $\Gamma_0[\hat{g}]\equiv \Gamma_0[g,\tau]$ which describes 
the non anomalous part of the renormalized action 
\beq
\Gamma_0[\hat{g}]\sim \sum_n \mathcal J_n[\hat{g}]\, .
\eeq
Here we recall the structure of the operator that are at most marginal from the Renormalization Group viewpoint.
The first term that can be included is trivial, corresponding to a cosmological constant contribution 
\beq
\mathcal S^{(0)}_\tau =  \Lambda^6\, \int d^6x\, \sqrt{\hat{g}}=  
\Lambda^6\, \int d^6x\, \sqrt{g}\,e^{-\frac{6\,\tau}{\Lambda}} \, .
\eeq
Here the superscript number in round brackets in $\mathcal{S}^{(n)}$ denotes the order of the contribution in the derivative expansion, so 
to distinguish the scaling behaviour of the various terms under the variation of the length scale.

For $n=1$ we obtain the operator  which reproduces the kinetic term of the dilaton  extensively discussed above.
Here we just mention that (\ref{KinTau}) can be derived from a general, d-dimensional formula 
for Weyl gauging the Einstein-Hilbert action,
\beq
\mathcal{S}_{\tau}^{(2)}= 
- \frac{\Lambda^{d-2}\, \left(d-2\right)}{8\,\left(d-1\right)}\, \int d^dx\, \sqrt{\hat g}\, \hat R = 
- \frac{\Lambda^{d-2}\, \left(d-2\right)}{8\,\left(d-1\right)}\, \int d^dx\, \sqrt{g}\, e^{\frac{(2-d)\,\tau}{\Lambda}}\, \bigg
[R - \left(d-1\right)\,\left(d-2\right)\, \frac{(\pd\tau)^2}{\Lambda^2} \bigg] \, ,
\label{one}
\eeq
which is exactly (\ref{KinTau}) for $d=6$.

The possible 4-derivative terms (n=2) are
\beq
\label{4der}
\int d^6x\, \sqrt{\hat g}\, \bigg( \alpha\, \hat R^{\mu\nu\rho\sigma}\, \hat R_{\mu\nu\rho\sigma} 
+ \beta\, \hat R^{\mu\nu}\, \hat R_{\mu\nu}+ \gamma\, \hat R^2 + \delta\, \hat \Box \hat R \bigg)\, .
\eeq
The $\square R$ contribution in this expression can be obviously omitted, being a total derivative.
We can also replace the Riemann tensor with the Weyl tensor  squared  (see (\ref{Weyl})) and remain with only two 
(as $\sqrt{\hat g}\,\hat C^{\mu\nu\rho\sigma}\, \hat C_{\mu\nu\rho\sigma}= 
\sqrt{g}\,C^{\mu\nu\rho\sigma}\, C_{\mu\nu\rho\sigma}\, e^{\frac{2\tau}{\Lambda}} $)
non trivial contributions, $\hat R^{\mu\nu}\, \hat R_{\mu\nu}$ and $\hat R^2$.
We present here the expression of (\ref{4der}) for a conformally flat metric, while the result for a general gravitational background 
can be computed exploiting the Weyl gauged tensors given in Appendix \ref{Geometrical2},
\bea
S^{(4)}_\tau 
&=& 
\int d^6x\, \sqrt{\hat g}\, \bigg( \alpha\, \hat R^{\mu\nu}\, \hat R_{\mu\nu} + \beta\, \hat R^2\bigg)
\nn \\
&=& 
\int d^6x\, e^{-\frac{2\,\tau}{\Lambda}}\, \bigg[ 
100\, \alpha\, \bigg( \frac{\Box\tau}{\Lambda} - 2\,\frac{\left(\pd\tau\right)^2}{\Lambda^2} \bigg)^2 
 + 2\, \beta\, \bigg( 15\,\frac{\left(\Box\tau\right)^2}{\Lambda^2}  
- \frac{68}{\Lambda^3}\, \Box\tau\, \left(\pd\tau\right)^2 + 72\, \frac{\left(\pd\tau\right)^4}{\Lambda^4}  \bigg)
\bigg]\, .
\eea
The last contributions that are significant down to the infrared regime are the marginal ones, i.e. the 6-derivative operators. 
To derive them we follow the analysis in \cite{Elvang:2012st}. We use the basis of diffeomorphic invariants of order 6 in the derivatives,
on which the $I_i's$ are expanded (see Eq. (\ref{WeylInv6})). It is made of 11 elements, 6 of which contain the Riemann tensor, that can 
be traded for a combination of the Weyl tensor and the Ricci tensor and scalar, so that we are left with only the 5 terms in 
$(K_1 - K_{11})$   (see Sec. (\ref{Basis})) that do not contain the Riemann tensor.
As we are going to write down the result only in the flat limit, we can exploit two additional constraints.
Indeed in \cite{Anselmi:1999uk} it was shown that, in this case, the integral of
\beq
R^3 - 11\, R\,R^{\mu\nu}\, R_{\mu\nu} + 30\,  {R_\mu}^\nu\,{R_\nu}^\alpha\,{R_\alpha}^\mu
- 6\, R\Box R + 20\, R^{\mu\nu}\Box R_{\mu\nu}
\eeq
vanishes, so that we can use this result to eliminate $R^{\mu\nu}\Box R_{\mu\nu}$. \\
Then, as the Euler density can be written in the form
\bea
E_6 
&=&
\frac{21}{100}\, R^3 - \frac{27}{20}\, R\, R^{\mu\nu}\, R_{\mu\nu}
+ \frac{3}{2}\, {R_\mu}^\nu\,{R_\nu}^\alpha\, {R_\alpha}^\mu
+ 4\, C_{\mu\nu\rho\sigma}\,{C^{\mu\nu}} _{\alpha\beta}\, C^{\rho\sigma\alpha\beta}
\nn \\
&&
-\, 8\, C_{\mu\nu\rho\sigma}\,C^{\mu\alpha\rho\beta}\, {{{C^\nu}_\alpha}^\sigma}_\beta
-6\, R_{\mu\nu}\, C^{\mu\alpha\rho\sigma}\, {C^\nu}_{\alpha\rho\sigma}
+\frac{6}{5}\, R\, C^{\mu\nu\rho\sigma}\, C_{\mu\nu\rho\sigma}
- 3\, R^{\mu\nu}\, R^{\rho\sigma}\, C_{\mu\rho\sigma\nu} \, .
\eea
It is apparent that only the first three terms are non vanishing on a conformally flat metric.
Now, as in the effective action these contributions are integrated and the Euler density is a total derivative,
one can thereby replace ${R_\mu}^\nu\,{R_\nu}^\alpha\,{R_\alpha}^\mu$ for $R^3$ and $R\,R^{\mu\nu}\,R_{\mu\nu}$.
In the end, Weyl gauging $R^3$, $R^{\mu\nu}\,R_{\mu\nu}$ and $R\Box R$ is sufficient to account for all the possible 
6-derivative terms of the dilaton effective action which do not vanish in the flat space limit.
After some integrations by parts, one can write the overall contribution as
\bea
S^{(6)}_\tau 
&=&
\int d^6x\, \sqrt{\hat{g}}\, \bigg[  \gamma\, \hat{R}^3 
+ \delta \hat{R}\,\hat R^{\mu\nu}\,\hat R_{\mu\nu}\, + \zeta\, \hat R \hat \Box \hat R  \bigg] \nn \\
&=&
\int d^6x\, 20\, \bigg[ \frac{1}{\Lambda^3}\,
\bigg( 5\,\zeta\, \Box^2\tau\, \Box\tau -(50\, \gamma + 7\, \delta - 30\, \zeta)\, (\Box\tau)^3 
- 8 \,(\delta + 5\,\zeta)\, \Box\tau\, (\pd\pd\tau)^2  \bigg)\nn \\
&& \hspace{14mm}
+\,  \frac{1}{\Lambda^4}\, \bigg( 50\, (6\,\gamma + \delta - 2\,\zeta)\, (\Box\tau)^2\, (\pd\tau)^2   
- 16\, (\delta + 5\, \zeta)\, \Box\tau\, \pd_\mu \pd_\nu\tau\, \pd^\mu\tau\, \pd^\nu\tau 
\nn \\
&& \hspace{14mm}
+\, 8\,(2\,\delta + 5\,\zeta)\, (\pd\tau)^2 (\pd\pd\tau)^2  \bigg)
- \frac{120}{\Lambda^5}\,(5\,\gamma + \delta - \zeta)\, \Box\tau\,(\pd\tau)^4 \nn \\
&& \hspace{14mm}
+   \frac{80}{\Lambda^6}\,(5\,\gamma + \delta - \zeta)\,(\pd\tau)^6 
\bigg].
\eea
We have introduced the compact notation $\left(\pd\tau\right)^n \equiv 
\left( \pd_\lambda\tau\,\pd^\lambda\tau \right)^{n/2}\, , \left(\pd\pd\tau\right)^2 \equiv 
\pd_\mu\pd_\nu\tau\, \pd^\mu\pd^\nu\tau $ to denote multiple derivatives of the dilaton field. 
The Weyl invariant part of the dilaton effective action is then given by
\beq
\Gamma_0[g,\tau] =  
\mathcal{S}^{(0)}_\tau + \mathcal{S}^{(2)}_\tau + \mathcal{S}^{(4)}_\tau + \mathcal{S}^{(6)}_\tau
+ \dots \, ,
\eeq
where the ellipsis denote all the possible higher-order, irrelevant terms.

\subsection{The counterterms and the anomaly} \label{Counterterms}

As we have discussed above, we construct the effective action by applying the Weyl gauging procedure to the renormalized effective action, 
which breaks scale invariance via the anomaly. First we must introduce the one-loop counterterm action, 
which is given, following \cite{Duff:1977ay} and \cite{Mazur:2001aa}, by the integrals of all the possible Weyl invariants
and of the Euler density continued to $d$ dimensions
\beq\label{CounterAction6}
{\Gamma}_{\textrm{Ct}}[g] = 
- \frac{\mu^{-\epsilon}}{\epsilon}\int d^d x\, \sqrt{g}\, \bigg( \sum_{i=1}^{3} c_i\, I_i + a E_6 \bigg) \, ,  
\quad \epsilon = 6 - d \, ,
\eeq
where $\mu$ is a regularization scale. It is this form of ${\Gamma}_{\textrm{Ct}}$, which is part of 
$\Gamma_{\textrm{ren}}$, to induce the anomaly relation 
\beq \label{AnomalyCondition}
\frac{2}{\sqrt{g}}g_{\mu\nu}\frac{\delta{\Gamma_{\textrm{ren}}}[g]}{\delta g_{\mu\nu}}\bigg|_{d\rightarrow 6} 
=\frac{2}{\sqrt{g}}g_{\mu\nu}\frac{\delta{\Gamma_{\textrm{Ct}}}[g]}{\delta g_{\mu\nu}}\bigg|_{d\rightarrow 6}= \mathcal{A}[g].
\eeq
In the derivation of the equation above, we have exploited the Weyl invariance of the non anomalous action $\Gamma_0[g]$ in $6$ dimensions
\beq \label{BareActionInvariance}
g_{\mu\nu}\frac{\delta{\Gamma}_0[g]}{\delta g_{\mu\nu}}\bigg|_{d\rightarrow 6} = 0,  \, 
\eeq
while the anomaly is generated entirely by the counterterm action ${\Gamma}_{\textrm{Ct}}[g]$, due to the relations 
\bea \label{NTracesCTI}
\frac{2}{\sqrt{g}}\, g_{\mu\nu}\, \frac{\delta}{\delta g_{\mu\nu}}\, \int d^d x\,\sqrt{g}\, I_i 
&=&
-\epsilon \, \bigg( I_i + \nabla_\mu J^\mu_i \bigg) \, , 
\\
\label{NTracesCTE}
\frac{2}{\sqrt{g}}\, g_{\mu\nu}\, \frac{\delta}{\delta g_{\mu\nu}}\, \int d^d x\, \sqrt{g}\, E_6
&=& 
-\epsilon \, E_6 \, ,
\eea
so that from (\ref{AnomalyCondition}) we find
\beq \label{TraceAnomaly6d}
\left\langle T \right\rangle = 
\frac{2}{\sqrt{g}}\, g_{\mu\nu}\, \frac{\delta{\Gamma}_{\textrm{Ct}}[g]}{\delta g_{\mu\nu}}\bigg|_{d\rightarrow 4} =  
\sum_{i=1}^3 c_i\, \left( I_i + \nabla_\mu J_i^\mu \right) + a E_6   \, .
\eeq
The explicit expressions of the derivative terms $\nabla_\mu J^\mu_i$ in Eq.~(\ref{NTracesCTI}) can be obtained using the functional 
variations listed in Appendix \ref{Geometrical3}. They are given by  
\bea
\nabla_\mu J_1^\mu 
&=& 
- \frac{3}{800}\, \nabla_\mu \bigg[
- 5\, \bigg( 44\, R^{\rho\sigma}\, \nabla^\mu R_{\rho\sigma} - 50\, R_{\rho\sigma}\, \nabla^\sigma R^{\mu\rho} 
- 3\, R^{\mu\nu}\, \pd_\nu R - 4\, R_{\nu\rho\sigma\alpha }\, \nabla^\mu R^{\nu\rho\sigma\alpha} 
\nn \\
&& \hspace{20mm}
+\, 40\, R^{\mu\rho\nu\sigma}\, \nabla_\nu R_{\rho\sigma} \bigg) + 19\, R\, \pd^\mu R  \bigg]
\nn \\
\nabla_\mu J_2^\mu 
&=& 
- \frac{3}{200}\, \nabla_\mu \bigg[
- 5\, \bigg( 4\, R^{\rho\sigma}\, \nabla^\mu R_{\rho\sigma} 
+ 10\, R_{\rho\sigma}\, \nabla^\sigma R^{\mu\rho} + 7\, R^{\mu\nu}\, \pd_\nu R 
- 4\, R_{\nu\rho\sigma\alpha }\, \nabla^\mu R^{\nu\rho\sigma\alpha}
\nn \\
&& \hspace{20mm}
-\, 40\, R^{\mu\rho\nu\sigma}\, \nabla_\nu R_{\rho\sigma}\,  \bigg) + 9 R\, \pd^\mu R  \bigg]
\nn \\
\nabla_\mu J_3^\mu
&=& 
\frac{1}{25}\, \nabla_\mu \bigg[ 
10\,  \bigg( 2\,\pd^\mu \square R - 5\, \nabla_\nu \square R^{\mu\nu} + R_{\nu\rho}\, \nabla^\mu R^{\nu\rho}
-2\, R^{\mu\nu}\, \pd_\nu R - R_{\nu\rho\sigma\alpha }\, \nabla^\mu R^{\nu\rho\sigma\alpha}
\nn \\
&& \hspace{15mm}
-\, 10\, R^{\mu\rho\nu\sigma}\, \nabla_\nu R_{\rho\sigma}\,  \bigg) - 3\, R\, \pd^\mu R  \bigg]\, .
\eea
The terms above are renormalization prescription dependent and are not present if, instead of the counterterms
$\sqrt{g}\, I_{i}$,  one chooses scalars  that are conformal invariant in $d$ dimensions, i.e. the $I^d_i$'s defined in 
Appendix \ref{Geometrical}. Notice that the inclusion of $d$-dimensional counterterms simplifies considerably the computation of the 
dilaton WZ action, as shown in \cite{Baume:2013ika}. In fact, in this scheme, the contribution of the $I^d_i$'s to the same action 
is just linear in the dilaton field and can be derived from the counterterm
\beq \label{IdCounterterm}
{\Gamma}^{d}_{\textrm{Ct}}[g] = - \frac{\mu^{-\epsilon}}{\epsilon}\, \int d^d x\, \sqrt{g}\, 
\bigg( \sum_{i=1}^{3} c_i\, I^d_i + a E_6 \bigg). 
\eeq
It can be explicitly checked that by expanding (\ref{IdCounterterm}) around $d=6$ and computing the order $O(\epsilon)$ contribution 
to the vev of the traced EMT one obtains the relation 
\beq \label{NTracesCT1}
\frac{2}{\sqrt{g}}\, g_{\mu\nu}\, \frac{\delta}{\delta g_{\mu\nu}}\, \int d^d x\,\sqrt{g}\, I^d_i = 
- \epsilon\, I_i + O\big( \epsilon^2 \big). \,
\eeq
In this simplified scheme it is possible to give the structure of the WZ action in any even dimension \cite{Baume:2013ika}, 
just by adding to the contribution of such invariants the one coming from the Euler density $E_d$, being the 
total derivative terms $\nabla_\mu J^\mu_i$ absent. 

\subsection{General scheme-dependence of the trace anomaly}

In this section we establish a connection between the two schemes used to derive the dilaton WZ action, with the inclusion of 
invariant counterterms of $B$ type which are either $d$ or 6-dimensional, 
in close analogy with the 4-dimensional case \cite{Coriano:2013xua}, that we now briefly review.

In this case one introduces the counterterm action \cite{Duff:1977ay}
\beq\label{CounterAction4}
{\Gamma}_{\textrm{Ct}}[g] = 
- \frac{\mu^{-\epsilon}}{\epsilon}\int d^d x\, \sqrt{g}\, \bigg( \beta_a\, F + \beta_b\, G\bigg) \, ,  \quad \epsilon = 4 - d \, ,
\eeq
where $\mu$ is a regularization scale. It is this form of ${\Gamma}_{\textrm{Ct}}$ to induce the anomaly condition 
\beq
\label{qt}
\frac{2}{\sqrt{g}}g_{\mu\nu}\frac{\delta{\Gamma_{\textrm{ren}}}[g]}{\delta g_{\mu\nu}}\bigg|_{d\rightarrow 4} = 
\frac{2}{\sqrt{g}}g_{\mu\nu}\frac{\delta{\Gamma_{\textrm{Ct}}}[g]}{\delta g_{\mu\nu}}\bigg|_{d\rightarrow 4}= 
\mathcal{A}[g],
\eeq
where $G$ is the Euler density in 4 dimensions,
\beq \label{Euler4d}
G = R^{\alpha\beta\gamma\delta}R_{\alpha\beta\gamma\delta} - 4\,R^{\alpha\beta}R_{\alpha\beta} + R^2\, ,
\eeq
whwreas $F$ is the squared Weyl tensor, which reads, for generic dimension,
\beq \label{SquaredWeyld}
F_d \equiv
C^{\alpha\beta\gamma\delta}C_{\alpha\beta\gamma\delta}=
R^{\alpha\beta\gamma\delta}R_{\alpha\beta\gamma\delta} 
- \frac{4}{d-2}R^{\alpha\beta}R_{\alpha\beta}+\frac{2}{(d-1)(d-2)}R^2.
\eeq
Its $d=4$ realization, called simply $F$, appears in the trace anomaly equation  (\ref{TraceAnomaly}).
From the well known relations
\bea \label{NTracesCTF}
\frac{2}{\sqrt{g}}\, g_{\mu\nu}\, \frac{\delta}{\delta g_{\mu\nu}}\, \int d^d x\,\sqrt{g}\, F 
&=&
-\epsilon \, \left(F - \frac{2}{3}\, \Box R\right)\, , 
\\
\label{NTracesCTG}
\frac{2}{\sqrt{g}}\, g_{\mu\nu}\, \frac{\delta}{\delta g_{\mu\nu}}\, \int d^d x\, \sqrt{g}\, G 
&=& 
-\epsilon \, G \, ,
\eea
it follows that the explicit form of the trace anomaly equation (\ref{qt}) is
\beq \label{TraceAnomaly4d}
\left\langle T \right\rangle = 
\frac{2}{\sqrt{g}}\, g_{\mu\nu}\, \frac{\delta{\Gamma}_{\textrm{Ct}}[g]}{\delta g_{\mu\nu}}\bigg|_{d\rightarrow 4} =  
\beta_a\, \left( F - \frac{2}{3}\, \Box R\right) + \beta_b\, G   \,  .
\eeq
The $\Box R$ term in Eq.~(\ref{NTracesCTF}) is prescription dependent and can be avoided if the 
$F$-counterterm is chosen to be conformal invariant in $d$ dimensions, i.e. using the square $F_d$ of the Weyl tensor in $d$ 
dimensions in (\ref{SquaredWeyld}),
\beq \label{FdCounterterm}
{\Gamma}^{d}_{\textrm{Ct}}[g] = - \frac{\mu^{-\epsilon}}{\epsilon}\, \int d^d x\, \sqrt{g}\, \bigg( \beta_a F_d + \beta_b G
\bigg)\, .
\eeq
In fact, expanding (\ref{FdCounterterm}) around $d=4$ and computing the $O(\epsilon)$ 
contribution to the vev of the traced EMT it is found that
\bea
\int d^d x\,\sqrt{g}\, F_d 
&=& \int d^d x\,\sqrt{g}\, \bigg[ F - \epsilon\, \bigg( R^{\alpha\beta}R_{\alpha\beta} - \frac{5}{18}\, R^2 \bigg)
+ O\big(\epsilon^2\big) \bigg] \, , \\
\frac{2}{3}\, \Box R 
&=&
\frac{2}{\sqrt{g}}\, g_{\mu\nu}\, \frac{\delta}{\delta g_{\mu\nu}}\, 
\int d^4 x\, \sqrt{g}\, \left( R^{\alpha\beta} R_{\alpha\beta} - \frac{5}{18} R^2 \right) \, .
\eea
These formulae, combined with (\ref{NTracesCTF}), give
\beq \label{NTracesCT2}
\frac{2}{\sqrt{g}}\, g_{\mu\nu}\, \frac{\delta}{\delta g_{\mu\nu}}\, \int d^d x\,\sqrt{g}\, F_d
= - \epsilon\, F + O\big(\epsilon^2\big) \,
\eeq
in which the $\square R$ term is now absent. \\

For $d=6$ we proceed in a similar way. We expand the $d$-dimensional counterterms around $d=6$ to identify the finite contributions as
\beq \label{SeriesEps}
I_i^d = I_i + (d-6)\, \frac{\pd I^d_i}{\pd d}\bigg|_{d=6} = I_i - \epsilon\, \frac{\pd I^d_i}{\pd d}\bigg|_{d=6}\, .
\eeq
Using (\ref{SeriesEps}) in the $d$-dimensional counterterms, we have
\beq
-\frac{1}{\epsilon}\, \int d^d x\, \sqrt{g}\, I^d_i = 
-\frac{1}{\epsilon} \int d^d x\, \sqrt{g}\, I_i + \int d^dx\, \sqrt{g}\, \frac{\pd I^d_i}{\pd d}\bigg|_{d=6}\, .
\eeq
This implies, due to (\ref{NTracesCTI}) and (\ref{NTracesCT1}), that 
\bea
-\frac{1}{\epsilon}\, \frac{2}{\sqrt{g}}\, g_{\mu\nu}\, \frac{\delta}{\delta g_{\mu\nu}}\, \int d^d x\,\sqrt{g}\, I_i 
&=&
I_i -  \frac{2}{\sqrt{g}}\, g_{\mu\nu}\, \frac{\delta}{\delta g_{\mu\nu}}\, \int d^dx\, \sqrt{g}\, \frac{\pd I^d_i}{\pd d}\bigg|_{d=6}
\eea
and hence 
\bea
\frac{2}{\sqrt{g}}\, g_{\mu\nu}\, \frac{\delta}{\delta g_{\mu\nu}}\, \int d^dx\, \sqrt{g}\, \frac{\pd I^d_i}{\pd d}\bigg|_{d=6}
&=& 
- \nabla_\mu J^\mu_i\, .
\eea
This clearly identifies the local counterterms that we can add to (\ref{CounterAction6})
in order to arbitrarily vary the coefficients $c_i$ in (\ref{TraceAnomaly6d}). They are given by
the derivatives of the $d$-dimensional terms $I_i$ evaluated at $d=6$, linearly combined with arbitrary coefficients $c'_i$
\bea
{\Gamma'}_{\textrm{Ct}}[g] = 
- \frac{\mu^{-\epsilon}}{\epsilon}\int d^d x\, \sqrt{g}\, \bigg( \sum_{i=1}^{3} c_i\, I_i + a E_6 \bigg)
+ \int d^6x\, \sqrt{g}\, \sum_{i=1}^3 c'_i\, \frac{\pd I_i^d}{\pd d}\bigg|_{d=6} \nn \\
\label{CounterAction6Mod}
\eea
which gives 
\bea
\left\langle T' \right\rangle \equiv
\frac{2}{\sqrt{g}}\, g_{\mu\nu}\, \frac{\delta{\Gamma'}_{\textrm{Ct}}[g]}{\delta g_{\mu\nu}}\bigg|_{d\rightarrow 4} =  
\sum_{i=1}^3 c_i\, I_i  + a E_6 + \sum_{i=1}^3 \left( c_i - c'_i \right)\, \nabla_\mu J_i^\mu   \, .
\label{CounterAction6Mod1}
\eea
The choice $c'_i=c_i$ in (\ref{CounterAction6Mod})  then allows to move back to the alternative scheme in which the local anomaly 
contribution is not present. 
We list the three local counterterms of (\ref{CounterAction6Mod}). They are given by 
\bea
\frac{\pd I^d_1}{\pd d}\bigg|_{d=6}
&=& 
\frac{1}{16000}\, \bigg(-307\, K_1 + 3465\, K_2 - 540\, K_3 - 3750\, K_4 + 6000\, K_5 + 3000\, K_6\bigg) \, , \nn \\
\frac{\pd I^d_2}{\pd d}\bigg|_{d=6} 
&=& 
\frac{1}{4000}\, \bigg( -167\, K_1 + 1965\, K_2 - 540\, K_3 - 2750\, K_4 + 3000\, K_5 + 3000\, K_6\bigg) \, , \nn \\
\frac{\pd I^d_3}{\pd d}\bigg|_{d=6} 
&=& 
\frac{1}{500}\, \bigg(-18\,K_1 + 140\,K_2 - 90\, K_3 - 70\,K_9 + 500\,K_{10} - 250\,K_{11} + 25\,K_{12} 
\nn \\ 
&& \hspace{11mm}
-\,  625\,K_{13} + 750\,K_{15}\bigg) \, .\nn\\
\eea
Finally, in general one might also be interested to generate an anomaly functional in which 
the derivative terms appear in combinations thata are different from those in the trace anomaly equation (\ref{anom6D}). 
For this goal, one should use proper linear combinations of the $K_i$ according to the relations listed in (\ref{Geometrical3}).

\subsection{Gauging 6-dimensional counterterms}

At this point we follow the same approach as in \cite{Coriano:2013xua} to Weyl-gauge the renormalized effective action.

We expand the gauged counterterms in a double power series with respect to $\epsilon = 6-d$
and $\kappa_{\Lambda}\equiv1/\Lambda$ around $(\epsilon , \kappa_\Lambda )=( 0,0)$. 
In their formal expansion
\beq \label{PreGauging}
- \frac{1}{\epsilon}\,\int d^dx \, \sqrt{\hat{g}} \, \hat{I}_i\, (\hat{E}_6) = 
- \frac{1}{\epsilon}\, \int d^dx \sum_{i,j=0}^{\infty}\frac{1}{i!j!}\, \epsilon^i\,\left( \kappa_\Lambda\right)^j\,
\frac{\pd^{i+j}\left(\sqrt{\hat g}\, \hat I_i\, (\hat E_6)\, \right)}{\pd \epsilon^i\,\pd \kappa_\Lambda^{j}}\, 
\eeq
only the $O(\epsilon)$ contributions are significant, due to the $1/\epsilon$ factor in front of the counterterms. 
On the other hand, similarly to the case in $4$ dimensions, the condition
\beq
\frac{\pd^{n} \left(\sqrt{\hat{g}} \, \hat I_i(\hat E_6)\right)}{\pd\kappa_\Lambda^{n}} = O(\epsilon^2) \, , 
\quad n \geq 7 \, 
\eeq
holds, as the Euler density and the three conformal invariants are at most cubic in the Riemann tensor
and in its double covariant derivatives. Besides, there are no terms with more than two dilatons in the gauged Riemann tensor 
(see Appendix \ref{Geometrical}). 
All the terms which are of $O(1/\epsilon)$ in (\ref{PreGauging}) and are different from $I_i\, (E_6)$
are found to vanish after some integrations by parts. 
So in general, after gauging the counterterms we end up with the general result
\beq
- \frac{\mu^{-\epsilon}}{\epsilon}\,\int d^dx \, \sqrt{\hat{g}} \, \hat{I}_i \,(\hat E_6 ) =
- \frac{\mu^{-\epsilon}}{\epsilon}\, \int d^dx \, \sqrt{g} \, I_i\,(E_6) +  \Sigma_{i(a)} + O(\epsilon)  \, .
\eeq
where the $\Sigma$'s are
\bea
&&
\Sigma_1 =  \int d^6 x \, \sqrt{g}\,  \bigg\{- \frac{\tau}{\Lambda}\, \bigg( I_1 + \nabla_\mu J_1^\mu \bigg) \nn \\ 
&&
+\, \frac{1}{\Lambda^2}\, \bigg[ \frac{3}{4}\, R^{\mu\rho\sigma\alpha}\, 
{R^\nu}_{\rho\sigma\alpha}\, \pd_\mu \tau\, \pd_\nu \tau
- \frac{3}{40}\, R^{\mu\nu\rho\sigma}\, R_{\mu\nu\rho\sigma}\, \left(\pd \tau\right)^2 
- \frac{3}{10}\, R\, \left(\nabla \pd \tau\right)^2 \nn \\ 
&& 
+\, \frac{9}{4}\, R^{\mu\rho\sigma\nu}\, R_{\rho\sigma}\, \pd_\mu \tau\, \pd_\nu \tau
- 3\, R^{\mu\nu\rho\sigma}\, \nabla_\nu \pd_\rho \tau\, \nabla_\mu \pd_\sigma \tau
-\frac{57}{800}\, R^2\, \left(\pd \tau\right)^2 \nn \\
&& 
-\, \frac{21}{16}\, R^{\mu\rho}\, {R_\rho}^\nu\, \pd_\mu \tau\, \pd_\nu \tau  
- \frac{9}{4}\, R^{\mu\nu}\, \Box \tau\,\nabla_\mu \pd_\nu \tau
+ \frac{57}{160}\, R^{\mu\nu}\,R_{\mu\nu}\, \left(\pd \tau\right)^2 \nn \\
&& 
+\, \frac{3}{2}\, R^{\mu\nu} \nabla^\rho \pd_\mu \tau \, \nabla_\rho \pd_\nu \tau 
+ \frac{57}{80}\, R\, R^{\mu\nu} \pd_\mu \tau \, \pd_\nu \tau 
+ \frac{57}{160}\, R\, \left(\Box\tau\right)^2  \bigg] \nn \\
&& 
+\, \frac{1}{\Lambda^3}\, \bigg[
- \frac{7}{16}\, \left(\Box \tau \right)^3
+ \frac{3}{2}\, \left(\nabla \pd \tau \right)^2 \, \Box \tau
- 6\, R^{\mu\nu\rho\sigma}\, \pd_\rho \tau\, \pd_\nu \tau\, \nabla_\mu \pd_\sigma \tau \nn \\ 
&& 
+ 3\, R^{\mu\nu}\, \nabla_\rho \pd_\nu \t \, \pd_\mu \t\, \pd^\rho \t 
-\frac{9}{4}\, R^{\mu\nu}\, \pd_\mu \tau \,\pd_\nu \tau \,\Box \tau
- \frac{3}{5}\, R\, \pd^\mu \tau\, \pd^\nu \tau\, \nabla_\mu \pd_\nu \tau \bigg] \nn \\
&& 
+\, \frac{1}{\Lambda^4}\, \bigg[
- \frac{3}{2}\, \left(\pd \t \right)^2\, \left(\nabla \pd \t \right)^2
-\, \frac{3}{8}\,  \left(\pd \t \right)^2\, \left(\Box \tau\right)^2
+ \frac{3}{4}\, \pd^\mu \left( \pd\tau \right)^2\, \pd_\mu \left( \pd\tau \right)^2
-\frac{3}{20}\, R\,\left(\pd \t\right)^4 
\bigg] \nn \\
&& 
+\, \frac{1}{\Lambda^5}\, \frac{3}{2}\, \left(\pd \t\right)^4\, \Box \t
- \frac{\left(\pd \t\right)^6}{\Lambda^6}\,     \bigg\},
\label{WZI1}
\eea
for the first
\bea
&&
\Sigma_2 = \int d^6 x \, \sqrt{g}\,  \bigg\{- \frac{\tau}{\Lambda}\, \bigg( I_2 + \nabla_\mu J_2^\mu \bigg) \nn \\ 
&& 
+ \frac{1}{\Lambda^2}\, \bigg[
3\, R^{\mu\rho\sigma\alpha}\, {R^\nu}_{\rho\sigma\alpha}\, \pd_\mu \tau\, \pd_\nu \tau + \frac{27}{40}\, R\,  \left(\Box\t\right)^2 
- \frac{6}{5}\,  R\, \left(\nabla \pd \t\right)^2  - \frac{27}{200}\, R^2\, \left(\pd \t \right)^2 \nn \\
&& 
-\, \frac{3}{10}\, R^{\mu\nu\rho\sigma}\, R_{\mu\nu\rho\sigma}\, \left(\pd  \tau\right)^2 
+ 3\, R^{\mu\rho\sigma\nu}\, R_{\rho\sigma}\, \pd_\mu \tau\, \pd_\nu \tau
- \frac{15}{4}\, R^{\mu\rho}\, {R_\rho}^\nu\, \pd_\mu \tau\, \pd_\nu\tau \nn \\ 
&& 
-\, 3\, R^{\mu\nu}\, \nabla_\nu \pd_\mu \tau\, \Box\tau + \frac{27}{40}\, R^{\mu\nu}\, R_{\mu\nu}\, \left(\pd\tau\right)^2
+ 6\, R^{\mu\nu}\, \nabla^\rho \pd_\mu \tau\, \nabla_\rho  \pd_\nu \tau
+\frac{27}{20}\, R\, R^{\mu\nu}\, \pd_\mu\tau\, \pd_\nu \tau  \bigg]  \nn \\ 
&& 
+\, \frac{1}{\Lambda^3}\, \bigg[ 
\frac{11}{4}\, \left(\Box \tau \right)^3 - 6\, \left(\nabla \pd \tau \right)^2\, \Box \tau
- 8\,  R^{\mu\nu}\, \nabla_\rho\pd_\nu \tau \, \pd_\mu\tau \, \pd^\rho \tau
- 6\,  R^{\mu\nu}\, \nabla_\mu \pd_\nu \tau \left(\pd\tau\right)^2
\nn \\
&& 
+\, 8\, R^{\mu\nu\rho\sigma}\, \pd_\nu\tau\, \pd_\rho\tau\, \nabla_\mu\pd_\sigma\tau 
+ 5\, R^{\mu\nu}\, \pd_\mu \tau \, \pd_\nu \tau\, \Box \tau
+ \frac{18}{5}\, R\, \pd^\mu \t\, \pd^\nu \t\, \nabla_\mu \pd_\nu \t  
+ 3\, R\,  \left(\pd \t\right)^2\, \Box \t \bigg]  \nn \\
&& 
+\, \frac{1}{\Lambda^4}\, \bigg[ 
6\, \left(\pd \t\right)^2\, \left(\nabla \pd \t\right)^2 - \frac{9}{2}\, \left(\pd \t\right)^2\, \left(\Box \t\right)^2 
- 3\, \pd^\mu \left( \pd\tau \right)^2\, \pd_\mu \left( \pd\tau \right)^2 - \frac{3}{5}\, R\, \left(\pd \t\right)^4
\bigg]  \nn \\ 
&& 
+\, \frac{6}{\Lambda^5}\, \left(\pd \t\right)^4\, \Box \t - \frac{4}{\Lambda^6}\, \left(\pd \t\right)^6
\bigg\}, 
\label{WZI2}
\eea
%
%
for the second and
\bea 
&&
\Sigma_3 =
\int d^6 x \, \sqrt{g}\,  \bigg\{ - \frac{\tau}{\Lambda}\, \bigg( I_3 + \nabla_\mu J_3^\mu \bigg) 
\nn \\
&& 
+\, \frac{1}{\Lambda^2}\, \bigg[ -\frac{3}{25} R^2 \left(\pd \t \right)^2  + \frac{13}{10}R^{\m \n}R\, \pd_{\m}\t \pd_{\n}\t
-\frac{2}{5}\, R^{\mu\nu\rho\sigma}\, R_{\mu\nu\rho\sigma}\, \left(\pd \t \right)^2 \nn \\ 
&&
+\frac{9}{10}\, R\,  \left(\Box \t\right)^2  - \frac{3}{10}\, R\, \pd^\mu \tau \,\Box \pd_\mu \tau  
- \frac{12}{5}\, R\, \left(\nabla \pd \t \right)^2 - 5\, R^{\mu\rho}\, {R_\rho}^\nu\, \pd_\mu \tau\, \pd_\nu \tau \nn \\ 
&& 
+ 7\, R^{\mu\nu}\, \nabla_\mu \pd_\nu \t\, \Box \t  - 9\, R^{\mu\nu}\, \pd_\mu \tau \,\Box \pd_\nu \tau\,
- \Box \tau \, \Box^2 \t + \frac{2}{5}\, R^{\mu\nu}\, R_{\mu\nu}\, \left(\pd \t \right)^2 \nn \\ 
&& 
+ 8\, R^{\mu\nu} \,\nabla_\rho \nabla_\nu \pd_\mu \tau\, \pd^\rho \tau  
+ 16\, R^{\mu\nu\rho\sigma}\,\nabla_\nu \pd_\rho \tau \, \nabla_\mu \pd_\sigma \tau  \bigg]
\nn \\
&&
+\, \frac{1}{\Lambda^3}\, \bigg[
2\, \left(\Box \tau \right)^3 - 8\, \left(\nabla \pd \tau \right)^2\, \Box\tau
- \frac{16}{5}\, R\,\pd^\mu \tau \pd^\nu \tau\,\nabla_\mu \pd_\nu \tau 
+ 8\, R^{\mu\nu}\, \pd_\mu \t\, \pd_\nu \t\, \Box \tau
\nn \\
&&
+\, 32\, R^{\mu\nu\rho\sigma}\,\pd_\nu \tau \, \pd_\rho \tau \, \nabla_\mu \pd_\sigma \tau
\bigg] \nn \\
&&
+\, \frac{1}{\Lambda^4}\, \bigg[
- 4\, \left(\pd \t\right)^2\, \left(\Box \t \right)^2  - 4\, \pd^\mu \left( \pd\tau \right)^2\, \pd_\mu \left( \pd\tau \right)^2
+ 16\, \left(\pd \t\right)^2\, \left(\nabla \pd \t\right)^2  - \frac{4}{5}\, R\, \left(\pd \t\right)^4 \bigg]
\bigg\}
\label{WZI3}
\eea
for the third invariant, while the contribution from the integrated Euler density is
\bea
&&
\Sigma_a =
\int d^6 x \, \sqrt{g}\,  \bigg\{- \frac{\tau}{\Lambda}\,  E_6 
+\, \frac{1}{\Lambda^2}\, \bigg[ 
12\, R^{\mu\rho\sigma\alpha}\, {R^\nu}_{\rho\sigma\alpha}\, \pd_\mu \tau\, \pd_\nu \tau   
- 3\,R^{\mu\nu\rho\sigma}\, R_{\mu\nu\rho\sigma}\,\left(\pd \t \right)^2  \nn \\
&&
+\, 24\, R^{\mu\rho\sigma\nu}\, R_{\rho\sigma}\, \pd_\mu \tau\, \pd_\nu \tau
+ 12\, R^{\mu\nu}\, R_{\mu\nu}\,\left(\pd \t  \right)^2 
- 24\,R^{\mu\rho}\,{R^\nu}_\rho\,\pd_\mu \t\, \pd_\nu \t 
+ 12\, R\, R^{\mu\nu}\, \pd_\mu \tau\, \pd_\nu \tau -  3\, R^2\,\left(\pd\tau \right)^2  \bigg] \nn \\
&& 
+\, \frac{1}{\Lambda^3}\, \bigg[ 16\, R^{\mu\nu\rho\sigma}\, \pd_\nu \tau \, \pd_\rho \tau\, \nabla_\mu \pd_\sigma \tau
- 16\, R^{\mu\nu}\,\nabla_\mu \pd_\nu \t\, \left(\pd \t\right)^2 
+ 32\, R^{\mu\nu}\,\nabla_\mu \pd_\rho \t\, \nabla^\rho \pd_\nu \tau \nn \\
&&
-\, 8\, R\, \pd^\mu \tau \pd^\nu \tau\, \nabla_\mu \pd_\nu \tau 
+   8\, R\, \left( \pd \tau \right)^2\, \Box \tau - 16\, R^{\mu\nu}\, \pd_\mu \tau\, \pd_\nu \tau \bigg] \nn \\
&&
+\, \frac{1}{\Lambda^4}\, \bigg[ 24\, \left( \pd \t \right)^2\, \left(\nabla \pd \t\right)^2 
- 24\, \left(\pd \t \right)^2\, \left(\Box \t \right)^2 - 6\, R\, \left(\pd \t \right)^4 \bigg]
+ \frac{36}{\Lambda^5}\, \Box \t\, \left(\pd \t \right)^4 - \frac{24}{\Lambda^6}\, \left(\pd \t\right)^6
\bigg\}. \, 
\label{WZE6}
\eea
The derivation of (\ref{WZI1})-(\ref{WZE6}) is very involved and we have used several integration by parts to simplify the previous 
expressions. The Wess-Zumino effective action is then obtained from (\ref{coc}) and, in a general gravitational background, 
it is just given by the combination of (\ref{WZI1})-(\ref{WZE6}) with the proper coefficients, up to a minus sign, i.e.
\beq
\label{WZfinal}
\Gamma_{WZ}[g,\tau]  = - \bigg( \sum_{i=1}^3 c_i\, \Sigma_i  + a\, \Sigma_a \bigg)\, .
\eeq
In the flat space-time limit $(g_{\mu\nu}\to \delta_{\mu\nu})$ 
there are obvious simplifications and this takes the form 
\bea
\Gamma_{WZ}[\delta,\tau] 
&=&
- \int d^6x\, \sqrt{g}\, \bigg\{ 
- \frac{c_3}{\Lambda^2}\, \Box \tau\, \Box^2\tau
+ \frac{1}{\Lambda^3}\, \bigg[ 
\bigg( - \frac{7}{16}\, c_1 + \frac{11}{4}\, c_2 + 2\, c_3 \bigg)\, \left(\Box\tau\right)^3 \nn \\
&&
+ \bigg( \frac{3}{2}\, c_1 - 6\, c_2 - 8\, c_3 \bigg)\, \left(\pd \pd \tau \right)^2\, \Box\tau \bigg]
+ \frac{1}{\Lambda^4}\, \bigg[ 
  \bigg( -\frac{3}{2}\, c_1 + 6\, c_2 + 16\, c_3 + 24\, a\bigg)\, \left(\pd\tau\right)^2\, \left(\pd\pd\tau\right)^2 
\nn \\
&&
-\, 
\bigg( \frac{3}{8}\, c_1 + \frac{9}{2}\, c_2 + 4\, c_3 + 24\, a \bigg)\, \left(\pd\tau\right)^2\, \left(\Box\tau\right)^2
+ \bigg( \frac{3}{4}\, c_1 - 3\, c_2 - 4\, c_3 \bigg)\, \pd^\mu\left(\pd\tau\right)^2\, \pd_\mu\left(\pd\tau\right)^2
\bigg] \nn \\
&&
\frac{1}{\Lambda^5}\, \bigg( \frac{3}{2}\, c_1 + 6\, c_2 + 36\, a \bigg)\, \left(\pd\tau\right)^4\, \Box\tau
- \frac{1}{\Lambda^6}\, \bigg( c_1 +4\, c_2 + 24\, a \bigg)\, \left(\pd\tau\right)^6
\bigg\}\, .
\label{Effective6dFlat}
\eea
 Having obtained the most general form for the Wess-Zumino action for conformal anomalies in 6 dimensions, 
we now turn to discuss one specific example in $d=6$, previously studied within the AdS/CFT correspondence. This provides an application of the results of the previous sections.

\subsection{The WZ action action for a free CFT: the $(2,0)$ tensor multiplet}

In this section we are going to determine the coefficients of the WZ action for the (2,0) tensor multiplet in $d=6$, which has been investigated 
in the past in the context of the $AdS_7/CFT_6$ holographic anomaly matching.

Free field realizations of CFT's are particularly useful in the analysis of the anomalies and their matching between theories 
in regimes of strong and weak coupling, allowing to relate free and interacting theories of these types. 
In this respect, the analysis of correlation functions which can be uniquely fixed by the symmetry is crucial in order to compute the anomaly for 
theories characterized by different field contents in general spacetime dimensions. This is the preliminary step in order to investigate the
matching with other realizations which share the same anomaly content. These are correlation functions which contain up to 3 EMT's and 
that can be determined uniquely, in any dimensions, modulo a set of 
coefficients, such as the number of fermions, scalars and/or spin 1, 
which can be fixed within a specific field theory realization \cite{Osborn:1993cr, Erdmenger:1996yc}

While in $d=4$ these correlation functions can be completely identified by considering a generic theory which combines free scalar, fermions and 
gauge fields \cite{Osborn:1993cr, Erdmenger:1996yc},\cite{Coriano:2012wp}, in $d$ dimensions scalars and fermions need to be accompanied not by a spin 1 (a one-form) but by  a $\kappa$ -form ($d=2 
\kappa +2$). In $d=6$ this is a 2-form, $B_{\mu\nu}$  \cite{Bastianelli:2000hi}.  

Coming to specific realizations and use of CFT's in $d=6$, we mention that, for instance, the dynamics of a single M5 brane is described by a 
free $\mathcal N=(2,0)$ tensor multiplet which contains 5 scalars, 2 Weyl fermions and a 2-form whose strength is anti-selfdual. For $N$ 
coincident $M5$ branes, at large $N$ values, the anomaly matching between the free field theories realizations and the interacting $(2,0)$ 
CFT's, investigated in the $AdS_7\times S^4$ supergravity description, has served as an interesting test of the correspondence between the $A$
 and $B$ parts of the anomalies in both theories \cite{Bastianelli:2000hi, Bastianelli:1999ab}. 
 
 We have summarized in Table 1 the coefficients 
of the WZ anomaly action in the case of a scalar, a fermion and a non-chiral $B_{\mu\nu}$  form, which are the fields appearing in the (2,2) CFT. Anomalies in the (2,0) and the (2,2) theories are related just by a factor 1/2, after neglecting the gravitational anomalies related to the imaginary parts of the (2,0) multiplet \cite{Bastianelli:2000hi}.  

We have extracted the anomaly coefficients in Table 1 from \cite{Bastianelli:2000hi}, having performed a redefinition of the third invariant $I_3$ 
in the structure of the anomaly functional (\ref{anom6D}). We choose to denote with $\tilde I_i, \tilde J_i$ and $\tilde c_i$ the anomaly 
operators and coefficients in \cite{Bastianelli:2000hi} 
\beq \label{Correspondence}
\tilde I_1 = I_1\, , \quad \tilde I_2 = I_2\, , \quad \tilde I_3 = 3\, I_3 + 8\, I_1 -2\, I_2 \, .
\eeq
Actually in \cite{Bastianelli:2000hi} the third conformal invariant is given by
\beq \label{I3tilde}
\tilde I_3 \equiv
C^{\alpha\gamma\rho\sigma} \bigg(
{\delta_\alpha}^\beta\, \Box - 4\, {R_\alpha}^\beta + \frac{6}{5}\, {\delta_\alpha}^\beta \,R \bigg)\, C_{\beta\gamma\rho\sigma}
+ \bigg( 8\, {\delta_\alpha}^\kappa\, {\delta_\beta}^\lambda - \frac{1}{2}\, g_{\alpha\beta}\, g^{\kappa\lambda} \bigg)\, 
\nabla_\kappa\nabla_\lambda C^{\alpha\gamma\rho\sigma}\, {C^\beta}_{\gamma\rho\sigma}\, ,
\eeq
which differs from our choice, reported in Appendix \ref{Geometrical}.
The relation in (\ref{Correspondence}) between the third invariant $\tilde{I}_3$ and $I_3$ can be derived
expanding (\ref{I3tilde}) on the basis of the $K$-scalars given in Sec. \ref{Basis} and comparing it to the third of (\ref{WeylInv6}).

In light of (\ref{Correspondence}), as the conformal anomalies depend only on the field content of the theory, i.e.
\beq \label{Matching}
\mathcal{A}[g]=\sum_{i=1}^3 c_i\, \left( I_i + \nabla_\mu J^\mu_i \right)=
\sum_{i=1}^3 \tilde c_i\, \left( \tilde I_i + \nabla_\mu \tilde J^\mu_i \right) \, ,
\eeq
by replacing (\ref{Correspondence}) on the r.h.s. of (\ref{Matching}), 
we conclude that the relations between the anomaly coefficients $\tilde c_i$ and $c_i$ are
\beq \label{MatchCoefficients}
c_1 = \tilde c_1 + 8\, \tilde c_3\, , \quad 
c_2 = \tilde c_2 - 2\, \tilde c_3\, , \quad 
c_3 = 3\, \tilde c_3 \, .
\eeq
The WZ action can be derived from Eq. (\ref{WZfinal}) by inserting the expressions of the $c_i$'s and $a$ extracted from Table 1.
These can be specialized to the scalar (S), fermion (F) and to the 2-form (B) cases, thereby generating via (\ref{Matching}) the 
corresponding anomaly functionals. For the $(2,0)$ tensor multiplet this is obtained from the relation
\beq
\mathcal A^{T}[g] = \frac{1}{2}\, \bigg( 10\, \mathcal A^{S}[g] + 2\, \mathcal A^{F}[g] + \mathcal A^{B}[g] \bigg).
\eeq

\begin{table}
\label{tt1}
$$
\begin{array}{|c|c|c|c|c|}\hline
I & c_1\times 7!\,(4\,\pi)^3 & c_2 \times 7!\,(4\,\pi)^3 & c_3 \times 7!\,(4\,\pi)^3 & a \times 7!\,(4\,\pi)^3
\\ \hline\hline
S & \frac{20}{3} & -\frac{7}{3} & 6 & -\frac{5}{72}
\\ \hline
F & \frac{64}{3} & -112 & 120  & -\frac{191}{72}
\\ \hline
B & - \frac{3688}{3} & -\frac{3458}{3} & 540 & -\frac{221}{4}
\\ \hline
T & -560 & -700 & 420 & -\frac{245}{8} 
\\
\hline
\end{array}
$$
\caption{Anomaly coefficients for a conformally coupled scalar (S), a Dirac Fermion (F), a 2-form field (B)
and the chiral $(2,0)$ tensor multiplet (T), to be normalized by an overall $1/ (7!\,(4\pi)^3)$}
\label{AnomalyCoeff}
\end{table}

\section{Dilaton interactions and constraints from $\Gamma_{WZ}$}

Having extracted the structure of the Wess-Zumino action and thus of the anomaly-related dilaton interactions via
the Weyl gauging of the effective action, we now follow a perturbative approach in the inverse conformal scale $\kappa_\Lambda=
\frac{1}{\Lambda}$.
We proceed with a Taylor expansion of the gauged metric, which is given by 
\beq \label{SeriesInG}
\hat{g}_{\mu\nu} = g_{\mu\nu}\, e^{-2\,\kappa_{\Lambda}\tau} =
\bigg(\d_{\mu\nu} + \kappa\, h_{\mu\nu} \bigg)\, e^{-2\,\kappa_{\Lambda}\tau} =
\bigg(\delta_{\mu\nu} + \kappa\, h_{\mu\nu} \bigg)\,
\sum_{n=0}^{\infty} \frac{(-2)^n}{n!}\,(\kappa_{\Lambda}\,\tau)^n \, ,
\eeq
where $\kappa$ is the gravitational coupling constant in $6$ dimensions and we are using Euclidean conventions.
As we are considering only the dilaton contributions, we focus on the functional expansion of the
renormalized and Weyl-gauged effective action $\hat\Gamma_{\textrm{ren}}[g,\tau]$ with respect to $\kappa_{\Lambda}$. This is 
easily done using the relation
\beq \label{CompositeDiff}
\frac{\pd\hat\Gamma_{\textrm{ren}}[g,\tau]}{\pd\kappa_{\Lambda}} = 
\int d^dx\, \frac{\delta\hat\Gamma_{\textrm{ren}}[g,\tau]}{\delta\hat{g}_{\mu\nu}(x)}
            \frac{\pd\hat{g}_{\mu\nu}(x)}{\pd\kappa_{\Lambda}}\, .
\eeq
Applying (\ref{CompositeDiff}) repeatedly and taking (\ref{SeriesInG}) into account, the perturbative series takes the form
\bea \label{Expansion}
\hat\Gamma_{\textrm{ren}}[g,\tau]
&=&
\Gamma_{\textrm{ren}}[g,\tau]
+\, \frac{1}{2!\,\Lambda^2}\, \int d^d \xu d^d \xd\, 
\frac{\delta^2\hat\Gamma_{\textrm{ren}}[g,\tau]}{\delta\hat{g}_{\muu\nuu}(\xu)\delta\hat{g}_{\mud\nud}(\xd)}
\frac{\pd \hat{g}_{\muu \nuu}(\xu)}{\pd \kappa_\Lambda}\frac{\pd \hat{g}_{\mud\nud}(\xd)}{\pd\kappa_\Lambda}
\nn \\
&& \hspace{-10mm}
+\, \frac{1}{3!\, \Lambda^3}\, \bigg(\int d^d x_1 d^d \xd d^d \xt\,
\frac{\delta^3\hat\Gamma_{\textrm{ren}}[g,\tau]}
{\delta\hat{g}_{\muu\nuu}(\xu)\delta\hat{g}_{\mud\nud}(\xd)\delta\hat{g}_{\mut\nut}(\xt)}
\frac{\pd \hat{g}_{\muu \nuu}(\xu)}{\pd \kappa_\Lambda}\frac{\pd \hat{g}_{\mud \nud}(\xd)}{\pd \kappa_\Lambda}
\frac{\pd \hat{g}_{\mut \nut}(\xt)}{\pd \kappa_\Lambda}
\nn \\
&& \hspace{-10mm}
+\, 3\, \int d^d \xu d^d \xd\, 
\frac{\delta^2\hat\Gamma_{\textrm{ren}}[g,\tau]}{\delta\hat{g}_{\muu\nuu}(\xu)\delta\hat{g}_{\mud\nud}(\xd)}
\frac{\pd^2 \hat{g}_{\muu\nuu}(\xu)}{\pd \kappa_\Lambda^2}\frac{\pd\hat{g}_{\mud\nud}(\xd)}{\pd \kappa_\Lambda} \bigg) +\ldots
\eea
As we are interested in the flat space limit of the dilaton action,
we write (\ref{Expansion}) by taking the limit of a conformally flat background metric 
$(\hat{g}_{\mu\nu}\rightarrow \hat \delta_{\mu\nu} \equiv\delta_{\mu\nu}\, e^{- 2\,\kappa_{\Lambda}\tau})$ obtaining
\bea \label{FinalExp}
\hat\Gamma_{\textrm{ren}}[\delta,\tau]
&=&
\Gamma_{\textrm{ren}}[\delta,\tau] + 
\frac{1}{2!\,\Lambda^2}\,  
\int d^d \xu d^d \xd\, \langle T(\xu) T(\xd)\rangle\, \tau(\xu)\tau(\xd)
\nn \\
&& 
-\, \frac{1}{3!\,\Lambda^3}\, \bigg[ \int d^d \xu d^d \xd d^d \xt\,\langle T(\xu) T(\xd) T(\xt)\rangle\, \tau(\xu)\tau(\xd)\tau(\xt)
\nn \\
&&
+\, 6\,\int d^d \xu d^d \xd\, \langle T(\xu) T(\xd)\rangle\, (\tau(\xu))^2\tau(\xd) \bigg] + \ldots \, ,
\eea
where we have used Eq. (\ref{NPF}) in the definition of the correlators of the EMT' s and the obvious relation
\beq \label{MetricDil}
\frac{\pd^n \hat{g}_{\mu\nu}(x)}{\pd \kappa_{\Lambda}^n}\bigg|_{g_{\mu\nu}=\delta_{\mu\nu},\kappa_{\Lambda} = 0} = 
\left(-2\right)^n \, \left(\tau(x)\right)^n\, \delta_{\mu\nu}\, .
\eeq
From (\ref{FinalExp}) one may identify the expression of the flat limit of the Wess-Zumino action
$\Gamma_{WZ}= \Gamma_{\textrm{ren}}[\delta,\tau] - \hat\Gamma_{\textrm{ren}}[\delta,\tau]$
written in terms of the trace n-point correlators of EMT's. 
This expression has to coincide with Eq. (\ref{Effective6dFlat}), and by comparing the dilaton vertices extracted from 
(\ref{FinalExp}) and (\ref{Effective6dFlat}) one can easily obtain some consistency conditions between the two forms of the vertices. In 
particular, 
in any even $d$ dimensions the first $d$ correlators follow rather directly from the expressions of the first $d$ dilaton interactions. These
are the only non vanishing ones. At the same time, any correlator of rank-n with $n\,>\,d$ can be evaluated by requiring that all the vertices with 
more than $d$ dilatons vanish identically, thus allowing to extract recursively all the EMT's Green functions of the corresponding rank.

We denote with $\mathcal I_n(\xu,\dots,x_n)$ the dilaton vertices obtained by the functional differentiations of 
the Wess-Zumino action, expressed in the coordinate space
\beq \label{FuncDiffWZ}
\mathcal I_{n}(\xu,\dots,x_n) = 
\frac{\delta^n \left(\hat\Gamma_{\textrm{ren}}[\delta,\tau]-\Gamma_{\textrm{ren}}[\delta,\tau]\right)}
{\delta\tau(\xu)\dots\delta\tau(x_n)} 
= -  \frac{\delta^n \Gamma_{WZ}[\delta,\tau]}{\delta\tau(\xu)\dots\delta\tau(x_n)}
\eeq
which can be promptly transformed to momentum space.
The expressions of such vertices up to the sixth order in $\kappa_{\Lambda}$ in momentum space are given by
\bea
{\mathcal I}_2(\ku,-\ku) 
&=& 
\kappa_{\Lambda}^2\, \langle T(\ku) T(-\ku)\rangle \, ,  
\nn \\
{\mathcal I}_3(\ku,\kd,\kt)
&=& 
-\kappa_{\Lambda}^3\, \bigg[ 
\langle T(\ku) T(\kd) T(\kt) \rangle +\, 2\, \sum_{i=1}^3 \langle T(k_i) T(-k_i)\rangle \bigg]\, ,
\nn \\
{\mathcal I}_4(\ku,\kd,\kt,\kq)
&=&
\kappa_{\Lambda}^4\, \Biggl[
\langle T(\ku) T(\kd) T(\kt) T(\kq) \rangle
+\, 2\,\sum_{\mathcal T\left\{4,(k_{i_1},k_{i_2})\right\}}
\langle T(k_{i_1}+k_{i_2}) T(k_{i_3}) T(k_{i_4})\rangle
\nn \\
&& \hspace{6mm}
+\, 4\, \bigg( \frac{1}{2}\, \sum_{\mathcal T\left\{4,(k_{i_1},k_{i_2})\right\}} 
\langle T(k_{i_1}+k_{i_2}) T(k_{i_3}+k_{i_4})\rangle
+ \sum_{i=1}^{4} \langle T(k_i) T(-k_i)\rangle \bigg) \Biggr]\, ,
\nn
\eea
\bea
{\mathcal I}_5(\ku,\kd,\kt,\kq,\kc) 
&=&
- \kappa_{\Lambda}^5\, \Biggl[
\langle T(\ku) T(\kd) T(\kt) T(\kq) T(\kc) \rangle 
\nn \\
&&
+\, 2\, \sum_{\mathcal T\left\{5,(k_{i_1},k_{i_2})\right\}}
\langle T(k_{i_1}+k_{i_2}) T(k_{i_3}) T(k_{i_4}) T(k_{i_5}) \rangle
\nn \\
&&
+ \,4\, \Biggl(
\sum_{\mathcal T\left\{5,(k_{i_1},k_{i_2},k_{i_3})\right\}} 
\langle T(k_{i_1}+k_{i_2}+k_{i_3}) T(k_{i_4}) T(k_{i_5}) \rangle
\nn \\
&&
+\, \sum_{\mathcal T\left\{5,[(k_{i_1},k_{i_2}),(k_{i_3},k_{i_4})]\right\}}
\langle T(k_{i_1}+k_{i_2}) T(k_{i_3}+k_{i_4}) T(k_{i_5}) \rangle \Biggr)
\nn \\
&&
+\, 8\, \Biggl(
\sum_{\mathcal T\left\{5,(k_{i_1},k_{i_2})\right\}} \langle T(k_{i_1}+k_{i_2}) T(-k_{i_1}-k_{i_2}) \rangle +
\sum_{i=1}^{5} \langle T(k_i) T(-k_i)\rangle \Biggr)
\Biggr]\, .
\nn \\
{\mathcal I}_6(\ku,\kd,\kt,\kq,\kc,\ks)
&=&
\kappa_{\Lambda}^6\, \Biggl[
\langle T(\ku) T(\kd) T(\kt) T(\kq) T(\kc) T(\ks) \rangle 
\nn \\
&&
+\, 2\, \sum_{\mathcal T\left\{6,(k_{i_1},k_{i_2})\right\}}
\langle T(k_{i_1}+k_{i_2}) T(k_{i_3}) T(k_{i_4}) T(k_{i_5}) T(k_{i_6}) \rangle
\nn \\
&&
+ \,4\, \Biggl( \sum_{\mathcal T\left\{6,(k_{i_1},k_{i_2},k_{i_3})\right\}}
\langle T(k_{i_1}+k_{i_2}+k_{i_3}) T(k_{i_4}) T(k_{i_5}) T(k_{i_6}) \rangle
\nn \\
&&
+ \, \sum_{\mathcal T\left\{6,[(k_{i_1},k_{i_2}),(k_{i_3},k_{i_4})]\right\}}
\langle T(k_{i_1}+k_{i_2}) T(k_{i_3}+k_{i_4}) T(k_{i_5}) T(k_{i_6}) \rangle 
\Biggr) \nn \\
&&
+\, 8\, \Biggl( 
\sum_{\mathcal T\left\{6,(k_{i_1},k_{i_2},k_{i_3},k_{i_4})\right\}}
\langle T(k_{i_1}+k_{i_2}+k_{i_3}+k_{i_4}) T(k_{i_5}) T(k_{i_6}) \rangle 
\nn \\
&&
+\, \sum_{\mathcal T\left\{6,[(k_{i_1},k_{i_2},k_{i_3}),(k_{i_4},k_{i_5})]\right\}}
\langle T(k_{i_1}+k_{i_2}+k_{i_3}) T(k_{i_4}+k_{i_5}) T(k_{i_6}) \rangle
\nn \\
&&
+\, \sum_{\mathcal T\left\{6,[(k_{i_1},k_{i_2}),(k_{i_3},k_{i_4})]\right\}}
\langle T(k_{i_1}+k_{i_2}) T(k_{i_3}+k_{i_4}) T((k_{i_5}+k_{i_6})) \rangle\Biggr)
\nn \\
&&
+\, 16\, \Biggl(
\frac{1}{2}\, \sum_{\mathcal T\left\{6,(k_{i_1},k_{i_2},k_{i_3})\right\}} 
\langle T(k_{i_1}+k_{i_2}+k_{i_3}) T(-k_{i_1}-k_{i_2}-k_{i_3}) \rangle
\nn \\
&&
+\, \, \sum_{\mathcal T\left\{6,(k_{i_1},k_{i_2})\right\}} \langle T(k_{i_1}+k_{i_2}) T(-k_{i_1}-k_{i_2}) \rangle
+ \sum_{i=1}^{6} \langle T(k_i) T(-k_i)\rangle
\Biggr)
\Biggr].
\label{DilIntStructure6}
\eea
These results can be easily extended to any higher order. 
The recipe, in this respect, is quite simple. To construct the vertex at order $n$ one has to sum to the $n$-point function
all the lower order functions in the hierarchy, down to $n=2$, partitioning the momenta in all the possible ways
and symmetrising each single contribution. The normalization factor in front of the correlator of order-$k$ is always $2^{n-k}$, 
while the factor in front of the 
vertex of order $n$ is $(-\kappa_\Lambda)^n$. Notice that, for $n$ even, we have an additional $1/2$ factor in front of the contributions from 
the two-point functions in which each EMT carries $n/2$ momenta, to avoid double counting.  
All the expressions in (\ref{DilIntStructure6}) have been thoroughly checked in 2 
dimensions, as illustrated in Appendix \ref{2D},  being their expression valid for any dimension.

We briefly recall the meaning of the notation used in (\ref{DilIntStructure6}) to organize the momenta.
The symbol $\mathcal T\left\{n,\dots\right\} $ is used to denote groups of momenta in the $n$-point function.
For example $\mathcal T\left\{4,(k_{i_1},k_{i_2})\right\}$ denotes, in the four point functions, the six possible pairs of distinct 
momenta.
\beq
\mathcal T\left\{4,(k_{i_1},k_{i_2})\right\} =
\left\{(\ku,\kd),(\ku,\kt),(\ku,\kq),(\kd,\kt),(\kd,\kq),(\kt,\kq) \right\} \, ,
\eeq
where we are combining the 4 momenta $k_1,...k_4$ into all the possible pairs, for a total of $\binom{4}{2}$ terms.
Moving to higher orders, the description of the momentum dependence is more complicated and we need to distribute the external 
momenta into several groups. For instance, the notation $\mathcal T\left\{5, [(k_{i_1},k_{i_2}),(k_{i_3},k_{i_4})]\right\}$ 
denotes the set of independent paired couples which can be generated out of $5$ momenta.
Their number is $15$ and they are given by
\bea
\mathcal T\left\{5,[(k_{i_1},k_{i_2}),(k_{i_3},k_{i_4})]\right\}
&=& 
\left\{[(\ku,\kd),(\kt,\kq)],[(\ku,\kd),(\kt,\kc)],[(\ku,\kd),(\kq,\kc)] 
\right. 
\nn \\
&& \hspace{-55mm}
\left.     
[(\ku,\kt),(\kd,\kq)],[(\ku,\kt),(\kd,\kc)],[(\ku,\kt),(\kq,\kc)],[(\ku,\kq),(\kd,\kt)],
[(\ku,\kq),(\kd,\kc)],[(\ku,\kq),(\kt,\kc)], 
\right.
\nn \\
&& \hspace{-55mm}    
\left.
[(\ku,\kc),(\kd,\kt)],[(\ku,\kc),(\kd,\kq)],[(\ku,\kc),(\kt,\kq)],[(\kd,\kt),(\kq,\kc)],
[(\kd,\kq),(\kt,\kc)],[(\kd,\kc),(\kt,\kq)] \right\}\, . 
\nn \\
\eea
At this point we move on to the evaluation of dilaton interactions and, consequently, of the first $6$ traced correlators, being clear from (\ref{DilIntStructure6}) that a direct computation of $\mathcal{I}_2 - \mathcal{I}_6$ from the anomaly action 
(\ref{Effective6dFlat}) allows to extract the structure of these Green functions.
Thus the dilaton interactions are
\bea \label{DilatonInt}
{\mathcal I}_2(\ku,-\ku) 
&=&
\frac{2}{\Lambda^2}\, c_3\, k_1^6\, ,  \nn \\
{\mathcal I}_3(\ku,\kd,\kt) 
&=& 
\frac{1}{\Lambda^3}\, \bigg[
\bigg( \frac{21}{8}\, c_1 - \frac{33}{2}\,c_2 - 12\,c_3 \bigg)\, k_1^2\,k_2^2\,k_3^2  \nn \\
&& \hspace{4mm}
+\, \bigg( - 3\, c_1 + 12\, c_2 + 16\,c_3 \bigg)\, \bigg( k_1^2\, \left(k_2\cdot k_3\right)^2 
+ k_2^2\, \left(k_1\cdot k_3\right)^2 + k_3^2\, \left(k_1\cdot k_2\right)^2 \bigg) \bigg] \, ,
\nn \\
{\mathcal I}_4(\ku,\kd,\kt,\kq)
&=&
\frac{1}{\Lambda^4}\, \bigg[
\bigg( 6\, c_1 - 24\,c_2 - 64\,c_3 - 96 \,a\bigg)\,
\sum_{\mathcal T\left\{4,(k_{i_1},k_{i_2})\right\}} k_{i_1}\cdot k_{i_2}\, \left(k_{i_3}\cdot k_{i_4}\right)^2  \nn \\
&& \hspace{-10mm}
+\, \bigg(\frac{3}{2}\, c_1 + 18\, c_2 + 16 \,c_3 + 96\,a \bigg)\,
\sum_{\mathcal T\left\{4,(k_{i_1},k_{i_2})\right\}} k_{i_1}\cdot k_{i_2}\, k_{i_3}^2\, k_{i_4}^2
\nn \\
&& \hspace{-10mm}
+\, \bigg( - 6\, c_1 + 24\,c_2 + 32\,c_3 \bigg)\,
\sum_{\mathcal T\left\{4,\left[(k_{i_1},k_{i_2}),(k_{i_3},k_{i_4})\right]\right\}} 
(k_{i_1}+k_{i_2})\cdot(k_{i_3}+k_{i_4})\, k_{i_1}\cdot k_{i_2}\, k_{i_3}\cdot k_{i_4} \bigg] \, ,
\nn
\eea
\bea
{\mathcal I}_5(\ku,\kd,\kt,\kq,\kc)
&=&
- \frac{12}{\Lambda^5}\, \bigg(c_1 + 4\,c_2 + 24\,a \bigg) \nn \\ 
&&
\times \sum_{\mathcal T\left\{5,(k_{i_1},k_{i_2},k_{i_3},k_{i_4})\right\}}
k_{i_5}^2\, \left( k_{i_1} \cdot k_{i_2} \, k_{i_3} \cdot k_{i_4} + k_{i_1} \cdot k_{i_3} \, k_{i_2} \cdot k_{i_4}
                                         + k_{i_1} \cdot k_{i_4} \, k_{i_2} \cdot k_{i_3} \right)\, , \nn \\
{\mathcal I}_6(\ku,\kd,\kt,\kq,\kc,\ks)
&=&
\frac{48}{\Lambda^6}\, \bigg(c_1 + 4\, c_2 + 24\,a\bigg)\, 
\sum_{\mathcal T\left\{6,\left[(k_{i_1},k_{i_2}),(k_{i_3},k_{i_4}),(k_{i_5},k_{i_6})\right]\right\}}
k_{i_1}\cdot k_{i_2}\, k_{i_3}\cdot k_{i_4}\, k_{i_5}\cdot k_{i_6} \, , \nn \\
\eea
(with $k_{i}^n \equiv (k_i^2)^{n/2}$).

These vertices can be used together with the relations (\ref{DilIntStructure6}) in order to extract the 
structure of the traced correlators. We find that the first two of them are given by
\bea \label{BuildingBlocks23}
\langle T(\ku) T(-\ku) \rangle 
&=& 
2\, c_3\, \ku^6 \, , 
\nn \\
\langle T(\ku) T(\kd) T(\kt) \rangle
&=& 
\bigg( 3\, c_1 - 12\,c_2 - 16\,c_3 \bigg)\, 
\bigg( \ku^2\,\left( \kd\cdot\kt \right)^2 + \kd^2\,\left( \ku\cdot\kt \right)^2 + \kt^2\,\left( \ku\cdot\kd \right)^2 \bigg)
\nn \\
&&
- \bigg( \frac{21}{8}\, c_1 - \frac{33}{2}\, c_2 - 12\, c_3 \bigg)\, \ku^2\,\kd^2\,\kt^2 
-\, 4\,c_3\,\bigg( \ku^6 + \kd^6 + \kt^6  \bigg) \, .
\eea
The structure of the four point Green function is much more complicated and is summarized in the expression
\bea
\langle T(\ku) T(\kd) T(\kt) T(\kq)\rangle 
&=&
\bigg[ 4\, c_3\, \bigg(7\, f^{2i,2i,2i} + 6\, f^{2i,2i,ij} + 3\, f^{2i,2i,2j}
+ 12\, f^{2i,ij,ij} + 12\, f^{2i,2j,ij} + 8\, f^{ij,ij,ij} \bigg) 
\nn \\
&&
+\, \bigg(-18\, c_1 + 72\, c_2 + 96\, c_3 \bigg)\, f^{2i,jk,jk}
+ 4\, \bigg(24\, a + 3\, c_1 - 12\, c_2 - 8\, c_3\bigg)\, f^{2i,2j,kl}
\nn \\
&&
-\, 6\, \bigg(16\, a + c_1 - 4\,c_2\bigg) f^{ij,kl,kl} 
+ \bigg(\frac{63}{4}\, c_1 - 99\, c_2 - 72\, c_3 \bigg)\, f^{2i,2j,2k}
\nn \\
&&
+\, \bigg(-6\,c_1 + 24\,c_2 + 32\,c_3 \bigg)\, \bigg( 2\, \, f^{2i,jk,jl} +  f^{ij,ik,jl} \bigg)
\bigg]\, .
\label{BuildingBlocks4}
\eea
Here we have introduced a compact notation for the basis of the 12 scalar functions $f^{\dots}(\ku,\kd,\kt,\kq)$ 
on which the correlator is expanded, leaving their dependence on the momenta implicit not to make the formula clumsy.
As each term in the Green function is necessarily made of three scalar products of momenta,
the role of the tree superscripts on each of the f scalars is to specify the way in which the momenta are distributed.
We present below the expressions of the first four scalar ${ f}$ 's, from which it should be clear how to derive the explicit forms of all the 
others. We obtain
\bea
f^{2i,2i,2i}(\ku,\kd,\kt,\kq)
&=&
\sum_{i=1}^{4} (k_i)^6 \, , \nn \\
f^{2i,2i,ij}(\ku,\kd,\kt,\kq)
&=&
\sum_{i=1}^{4} (k_i)^4\, \sum_{j\neq i} k_i \cdot k_j\, , \nn \\
\eea
\bea
f^{2i,2i,2j}(\ku,\kd,\kt,\kq)
&=&
\sum_{i=1}^{4} (k_i)^4\, \sum_{j\neq i} k_j^2\, , \nn \\
f^{2i,ij,ij}(\ku,\kd,\kt,\kq)
&=&
\sum_{i=1}^{4} (k_i)^2\, \sum_{j\neq i} \left(k_i \cdot k_j\right)^2\, .
\eea
Notice that each $f$-scalar is completely symmetric with respect to any permutation of the momenta, as for the whole correlator. The structure of 
the 5 and 6-point functions is similar to (\ref{BuildingBlocks23}), although they require broader bases of 
scalar functions to account for all their terms and we do not report them explicitly.

It is clear that the hierarchy in Eq. (\ref{hier}) can be entirely re-expressed in terms of the first six 
traced correlators. In fact, one notices that $\Gamma_{WZ}[\hat{\delta}]$ is at most of order 6 in $\tau$, with 
\beq \label{NoInt}
\mathcal I_n(\xu,\dots,x_n) = 0\, , \quad n \geq 7\, .
\eeq
Therefore, for instance, the absence of vertices with 7 dilaton external lines, which sets $\mathcal{I}_7=0$,
combined with the first 6 fundamental Green functions, are sufficient to completely fix the structure of the 7-point 
function, and so forth for the vertices of higher orders. In this way one can determine all the others recursively, up to the desired order.
The consistency of these relations could be checked, in principle, by a direct comparison with their expression obtained directly
from the hierarchy (\ref{hier}). This requires the explicit computation of functional derivatives of the anomaly functional $\mathcal{A}$ up to the relevant order, which is a much more time-consuming task.

\section{Conclusions}

We have presented a derivation of the Wess-Zumino conformal anomaly action using the Weyl gauging of the anomaly counterterms 
in $d=6$, previously discussed by us in the case of $d=4$. We have focused our attention on the contributions related to the local part of the anomaly.  
This result adds full generality to the analysis of dilaton effective actions, which carry an intrinsic 
regularization scheme dependence, due to the appearance in the anomaly functional of extra (total derivative) terms.
In general, the extraction of these extra contributions, as one can figure out from our study, is very involved, with a level of difficulty that grows 
with the dimensionality of the space in which the underlying CFT is formulated. Our main result for the WZ dilaton action in $d=6$ 
takes a simple form in flat space (\ref{Effective6dFlat}), and is characterized by 4 independent parameters $c_i$ and $a$, which 
express the field (particle) content of a certain conformal realization. Comparing our results with those of the previous literature 
\cite{Bastianelli:1999ab}, we have given the form of the WZ action in the case of the CFT of the $(2,0)$ tensor multiplet, which, in the past, has found application in the $AdS_7/CFT_6$ correspondence. 
We have also shown, in a second part of our analysis, using the structure of the dilaton action, that multiple correlators of traces of the EMT, for a 
theory with a certain anomaly content, are functionally related to correlation functions up to the sixth order and we have presented 
their explicit expressions up to rank-4.

\vspace{2cm}

\centerline{\bf \large Acknowledgements}

We thank Antonio Mariano for discussions. 

\appendix %

\section{Technical results}\label{Geometrical}



The definition of the Fourier transform for Eq. (\ref{NPF}) as well as for any n-point correlator is given by
\beq
\int \, d^d\xu\, \dots d^d x_n\, \left\langle T^{\muu\nuu}(\xu)\dots T^{\mu_n\nu_n}(x_n)\right\rangle \,
e^{-i(\ku\cdot \xu + \dots + k_n \cdot x_n)} = 
(2\pi)^d\,\delta^{(d)}\left( \sum_{i=1}^n k_i \right)\,\left\langle T^{\muu\nuu}(\ku)\dots T^{\mu_n\nu_n}(k_n)\right\rangle \, 
\label{NPFMom}
\eeq
with all the momenta in the vertex taken as incoming.
For the Riemann tensor we choose to adopt the sign convention
\bea \label{Tensors}
{R^\lambda}_{\mu\kappa\nu}
&=&
\pd_\nu \Gamma^\lambda_{\mu\kappa} - \pd_\kappa \Gamma^\lambda_{\mu\nu}
+ \Gamma^\lambda_{\nu\eta}\Gamma^\eta_{\mu\kappa} - \Gamma^\lambda_{\kappa\eta}\Gamma^\eta_{\mu\nu}\, .
\eea
The traceless part of the Riemann tensor in $d$ dimensions is the Weyl tensor
\beq \label{Weyl}
C_{\alpha\beta\gamma\delta} = R_{\alpha\beta\gamma\delta} -
\frac{1}{d-2}( g_{\alpha\gamma} \, R_{\beta\delta} - g_{\alpha\delta} \, R_{\beta\gamma}
             - g_{\beta\gamma}  \, R_{\alpha\delta} + g_{\beta\delta} \, R_{\alpha\gamma} ) +
\frac{R}{(d-1)(d-2)} \, ( g_{\alpha\gamma} \, g_{\beta\delta} - g_{\alpha\delta} \, g_{\beta\gamma})\, .
\eeq
It is also customary to introduce the Cotton tensor,
\beq \label{Cotton}
\tilde{C}_{\alpha\beta\gamma} = \nabla_{\gamma} K_{\alpha\beta} - \nabla_{\beta} K_{\alpha\gamma}\, , 
\quad \text{where}\, \quad
K_{\alpha\beta} = \frac {1}{d-2}\, \bigg( R_{\alpha\beta} - \frac{g_{\alpha\beta}}{2\,(d-1)}\, R \bigg)\, .
\eeq
The Weyl variations of the Christoffel symbols are
\bea \label{deltaWeylChristoffel}
\delta_{W} \Gamma^\alpha_{\beta\gamma}
&=&
- g_{\beta\gamma}\, \pd^\alpha \sigma + {\delta_\beta}^\alpha\, \pd_\gamma \sigma + {\delta_\gamma}^\alpha\, \pd_\beta\sigma
\quad \Rightarrow \quad \delta_{W} \Gamma^\alpha_{\alpha\gamma} = d\, \pd_\gamma \sigma \, , \nn \\
\nabla_\rho \delta_{W} \Gamma^\alpha_{\beta\gamma}
&=&
- g_{\beta\gamma}\, \nabla_\rho\pd^\alpha\sigma + {\delta_\beta}^\alpha\, \nabla_\rho\pd_\gamma\sigma
+ {\delta_\gamma}^\alpha\, \nabla_\rho\pd_\beta\sigma
\quad \Rightarrow \quad \delta_{W} \nabla_{\rho}\Gamma^\alpha_{\alpha\gamma} = d\, \nabla_\rho \pd_\gamma\sigma \, ,
\eea
which, using the Palatini identity
\beq \label{Palatini}
\delta {R^\alpha}_{\beta\gamma\rho} =
\nabla_{\rho}(\delta\Gamma^\alpha_{\beta\gamma}) - \nabla_{\gamma}(\delta\Gamma^\alpha_{\beta\rho})
\quad \Rightarrow \quad
\delta R_{\beta\rho} =
\nabla_{\rho} (\delta\Gamma^\lambda_{\beta\lambda}) - \nabla_{\lambda}(\delta\Gamma^\lambda_{\beta\rho})
\eeq
give the following Weyl variations of the Riemann and Ricci tensors
\bea \label{deltaWeylRiemann}
\delta_{W} {R^\alpha}_{\beta\gamma\rho} 
&=& 
  g_{\beta\rho}\, \nabla_{\gamma}\pd^\alpha\sigma 
- g_{\beta\gamma}\, \nabla_{\rho}\pd^\alpha \sigma
+ {\delta_{\gamma}}^{\alpha}\, \nabla_{\rho}\pd_\beta\sigma 
- {\delta_{\rho}}^{\alpha}\, \nabla_{\gamma}\pd_\beta\sigma \, , \nn \\
\delta_{W} R_{\beta\rho} 
&=& 
g_{\beta\rho}\, \Box \sigma + (d-2)\, \nabla_{\rho}\pd_\beta\sigma \, .
\eea
Is is also easy to use (\ref{deltaWeylChristoffel})-(\ref{deltaWeylRiemann}) 
to show that the variation of the Cotton tensor is simply given by
\beq \label{deltaWeylCotton}
\delta_{W} \tilde{C}_{\alpha\beta\gamma} = - \pd_\lambda\sigma\, {C^\lambda}_{\alpha\beta\gamma}\, ,
\eeq
which is expressed in terms of the Weyl tensor.

\subsection{Weyl scalars and the Euler density}\label{Geometrical2}

The three dimension-6 scalars that are Weyl invariant when multiplied by $\sqrt{g}$ and available in $d$ dimensions are 
%
\bea \label{WeylInvd}
I^d_1 &\equiv& 
C_{\mu\nu\alpha\beta}\, C^{\mu\rho\sigma\beta}\, {{C^\nu}_{\rho\sigma}}^\alpha  =
\frac{d^2+d-4}{(d-1)^2\,(d-2)^3}\, K_1 - \frac{3\,(d^2+d-4)}{(d-1)\,(d-2)^3}\, K_2 + \frac{3}{2\,d^2-6\,d+4}\, K_3 
\nn \\
&& \hspace{35mm}
+\, \frac{6\,d-8}{(d-2)^3}\, K_4 - \frac{3\,d}{(d-2)^2}\, K_5 - \frac{3}{d-2}\, K_6 + K_8 
\nn \\
I^d_2 &\equiv& 
C_{\mu\nu\alpha\beta}\, C^{\alpha\beta\rho\sigma}\, {C^{\mu\nu}}_{\rho\sigma} =
\frac{8\,(2\,d-3)}{(d-1)^2\,(d-2)^3}\, K_1 + \frac{72-48\,d}{(d-1)\,(d-2)^3}\, K_2 + \frac{6}{d^2-3\,d+2}\, K3
\nn \\
&& \hspace{35mm}
+\, \frac{16\,(d-1)}{(d-2)^2}\, K_4 - \frac{24}{(d-2)^2}\, K_5 - \frac{12}{d-2}\, K_6 + K_7
\nn \\
I^d_3
&\equiv&
\frac{d-10}{d-2}\, \bigg( \nabla^{\alpha}C^{\beta\gamma\rho\sigma}\, \nabla_{\alpha}C_{\beta\gamma\rho\sigma} 
- 4\, (d-2)\, \tilde{C}^{\gamma\rho\sigma}\, \tilde{C}_{\gamma\rho\sigma}\bigg) +
\frac{4}{d-2}\, \bigg(\square + \frac{2}{(d-1)}\, R\bigg)\,C^{\alpha\beta\rho\sigma}\,C_{\alpha\beta\rho\sigma}
\nn \\ 
&=&
\frac{16}{(d^2-3d+2)^2}\, K_{1} - \frac{32}{(d-1)\,(d-2)^2}\, K_{2} + \frac{8}{d^2-3\,d+2} K_{3} + \frac{16}{(d-1)\,(d-2)^2}\, K_{9}
\nn \\
&& \hspace{-5mm}
-\, \frac{32}{(d-2)^2}\, K_{10} + \frac{8}{d-2}\, K_{11} + \frac{4\,(d-6)}{(d-1)\,(d-2)^2}\, K_{12} 
+ \frac{88-12\,d}{(d-2)^2}\, K_{13} + K_{14} + \frac{8\,(d-10)}{(d-2)^2}\, K_{15}\, . \nn \\
\eea
In the gauging of the counterterms we use the following relations 
\bea
{\hat{\Gamma}}^\alpha_{\beta\gamma} 
&=& 
\Gamma^\alpha_{\beta\gamma} + \frac{1}{\Lambda}\, \bigg( {\delta_\beta}^\alpha\, \nabla_\gamma\tau 
+ {\delta_\gamma}^\alpha\, \nabla_\beta\tau - g_{\beta\gamma}\, \nabla^\alpha\tau \bigg) \, ,
\nn \\
\hat{R^\mu}_{\nu\rho\sigma}
&=& 
{R^\mu}_{\nu\rho\sigma}
+ g_{\nu\rho}\,   \bigg( \frac{\nabla_{\sigma}\pd^\mu\tau}{\Lambda}  + \frac{\pd^\mu\tau\, \pd_\sigma\tau}{\Lambda^2} \bigg)
- g_{\nu\sigma}\, \bigg( \frac{\nabla_{\rho}\pd^\mu\tau}{\Lambda}    + \frac{\pd^\mu\tau\, \pd_\rho\tau}{\Lambda^2}   \bigg)
\nn \\
&& +\, 
{\delta^\mu}_\sigma\,\bigg( \frac{\nabla_{\rho}\pd_\nu\tau}{\Lambda}   + \frac{\pd_\nu\tau\,\pd_\rho\tau}{\Lambda^2} \bigg) -
{\delta^\mu}_\rho\,  \bigg( \frac{\nabla_{\sigma}\pd_\nu\tau}{\Lambda} + \frac{\pd_\nu\tau\,\pd_\sigma\tau}{\Lambda^2} \bigg) +
\bigg( {\delta^\mu}_\rho\, g_{\nu\sigma} - {\delta^\mu}_\sigma\, g_{\nu\rho} \bigg)\,
\frac{(\pd\tau)^2}{\Lambda^2} \, , 
\nn \\
\hat R_{\mu\nu}
&=& 
R_{\mu\nu} - g_{\mu\nu}\, \bigg( \frac{\Box\tau}{\Lambda} - (d-2)\,\frac{(\pd\tau)^2}{\Lambda^2}\bigg)
- (d-2)\, \bigg( \frac{\nabla_\mu \pd_\nu\tau}{\Lambda} + \frac{\pd_\mu\tau\,\pd_\nu\tau}{\Lambda^2} \bigg)\, ,
\nn \\
\hat R
&\equiv&
\hat g^{\mu\nu}\, \hat R_{\mu\nu} =
e^{\frac{2\,\tau}{\Lambda}}\bigg[ R - 2\, (d-1)\, \frac{\Box \tau}{\Lambda} 
+ (d-1)\,(d-2)\, \frac{(\pd\tau)^2}{\Lambda^2} \bigg]\, .
\label{GaugeRiemann}
\eea
%
\subsection{Functional variations}\label{Geometrical3}

The results for the trace anomaly given in section \ref{Counterterms},
are obtained by computing the functional variations of the integrals of the $K_i$ in Dimensional Regularization.
A simple counting of the metric tensors needed to contract all the indices for any $K_i$ shows that
\beq
\delta_{W} \int d^dx\, \sqrt{g}\, K_i = \int d^dx\, \sqrt{g}\, \left[- \epsilon\, K_i + D(K_i) \right]\, \sigma
\, , \quad \epsilon = 6-d\, ,
\eeq
where the second term on the right hand side, $D(K_i)$, is a total derivative contribution.
We give the complete list of these terms below. We obtain
\bea \label{FuncVar}
D(K_1) &=& 12\, (d-1)\,\nabla_\mu \left(R\,\pd^\mu R \right) \nn \\
D(K_2) &=& \nabla_\mu \bigg[ 4\,(d-1)\, R_{\nu\rho}\, \nabla^\mu R^{\nu\rho}
           + 2\,\left(d-2\right)\, R^{\mu\nu}\, \pd_\nu R + \left(d+2\right)\, R\,\pd^\mu R \bigg]\nn \\
D(K_3) &=& 4\, \nabla_\mu \bigg[ R\, \pd^\mu R + 2\, R^{\mu\nu}\, \pd_\nu R
           +\left(d-1\right)\, R_{\nu\rho\sigma\alpha}\, \nabla^\mu R^{\nu\rho\sigma\alpha} \bigg] \nn \\
D(K_4) &=& 3\, \nabla_\mu \bigg[ \frac{d-2}{2}\,R^{\mu\nu}\, \pd_\nu R 
              +\left(d-2\right) R_{\nu\rho}\, \nabla^\rho R^{\mu\nu}
              +2\, R_{\nu\rho}\, \nabla^\mu R^{\nu\rho} \bigg] \nn \\
D(K_5) &=& \nabla_\mu \bigg[ -R\, \pd^\mu R - R^{\mu\nu}\, \pd_\nu R
           +2\, \left(d-1\right)\, R_{\nu\rho}\, \nabla^\rho R^{\mu\nu}
           -2\,d\, R_{\nu\rho}\,\nabla^\mu R^{\nu\rho} 
           + 2\,\left(d-2\right)\, R^{\mu\rho\nu\sigma}\, \nabla_\sigma R^{\nu\rho} \bigg] \nn \\
D(K_6) &=& \nabla_\mu \bigg[ 2\, R^{\mu\nu}\, \pd_\nu R + 4\, R_{\nu\rho}\, \nabla^\mu R^{\nu\rho}
           +\frac{d+2}{2}\, R_{\nu\rho\sigma\alpha}\, \nabla^\mu R^{\nu\rho\sigma\alpha}
           - 2\,d\, R^{\mu\rho\nu\sigma}\, \nabla_\sigma R_{\nu\rho} \bigg] \nn \\
D(K_7) &=& 6\, \nabla_\mu \bigg[ R_{\nu\rho\sigma\alpha}\, \nabla^\mu R^{\nu\rho\sigma\alpha} 
                            +4\, R^{\nu\rho\sigma\mu}\, \nabla_\rho R_{\nu\sigma}      \bigg] \nn \\
D(K_8) &=& 3\, \nabla_\mu \bigg[ \frac{1}{2}\,R_{\nu\rho\sigma\alpha}\, \nabla^\mu R^{\nu\rho\sigma\alpha}
                                 + 2\, R^{\nu\rho}\, \left( \nabla_\rho R_{\mu\nu} - \nabla_\mu R_{\nu\rho} \right)  \bigg] \nn \\
D(K_9) &=& \nabla_\mu \bigg[ 4\,\left(d-1\right)\, \pd^\mu \square R - \left(d-2\right)\, R\,\pd^\mu R \bigg] \nn \\
D(K_{10}) &=& \nabla_\mu \bigg[ 2\,\pd^\mu \square R + 2\,\left(d-2\right)\,\nabla_\nu \square R^{\mu\nu}
                               +2\, R^{\mu\nu}\, \pd_\nu R -4\, R_{\nu\rho}\, \nabla^\rho R^{\mu\nu} 
                               -\left(d-2\right)\, R_{\nu\rho}\,\nabla^\mu R^{\nu\rho}  \bigg]  \nn \\
D(K_{11}) &=& \nabla_\mu \bigg[ 8\, \nabla_\nu \square R^{\mu\nu} 
             - \left(d+2\right)\, R_{\nu\rho\sigma\alpha}\, \nabla^{\mu}R^{\nu\rho\sigma\alpha}
             - 16\, R^{\mu\rho\nu\sigma}\,\nabla_\sigma R_{\nu\rho} \bigg] \nn \\
D(K_{12}) &=& 4\, \nabla_\mu \bigg[ R\,\pd^\mu R - \left(d-1\right)\, \pd^\mu \square R \bigg]  \nn \\
D(K_{13}) &=& 2\,\ \nabla_\mu \bigg[ - \pd^\mu\square R  - \left(d-2\right)\, \nabla_\nu \square R^{\mu\nu} 
             -\, R^{\mu\nu}\, \pd_\nu R + 2\, R_{\nu\rho}\, \nabla^{\rho} R^{\mu\nu} 
             + 2\, R_{\nu\rho}\, \nabla^\mu R^{\nu\rho}   \bigg] \nn \\
D(K_{14}) &=& 8\, \nabla_\mu \bigg[ - \nabla_\nu \square R^{\mu\nu} 
              +R_{\nu\rho\sigma\alpha}\, \nabla^{\mu} R^{\nu\rho\sigma\alpha} 
              + 2\, R^{\mu\rho\nu\sigma}\,\nabla_\sigma R_{\nu\rho}  \bigg] \nn \\
D(K_{15}) &=& \nabla_\mu \bigg[ - \pd^\mu\square R - 2\,\left(d-2\right)\, \nabla_\nu \square R^{\mu\nu}
              -3\, R^{\mu\nu}\, \pd_\nu R + 6\, R_{\nu\rho}\, \nabla^\rho R^{\mu\nu} \nn \\
&& \hspace{10mm}
+ 2\, R_{\nu\rho}\, \nabla^\mu R^{\nu\rho} -2\, \left(d-2\right)\, R^{\mu\rho\nu\sigma}\,\nabla_\sigma R_{\nu\rho} \, \bigg]\, .
\eea
%

\section{The cases $d=2$ and $d=4$ as a check of the recursive formulae}\label{2D}
We briefly recall here how to cross-check the relations (\ref{DilIntStructure6}).

It is clear that the expressions of the dilaton vertices $\mathcal{I}_n$ given in (\ref{DilIntStructure6}) do not depend on the working dimensions. Therefore we take $d=2$ and check the agreement between the correlators that result from (\ref{DilIntStructure6}) 
and those found by a direct functional differentiation of the anomaly via the hierarchy (\ref{hier}). This provides a strong
check of the correctness of (\ref{DilIntStructure6}). In fact, the equation of the trace anomaly in 2 dimensions takes the form
\beq \label{TraceAnomaly2D}
\langle T \rangle = - \frac{c}{24\,\pi}\, R\, ,
\eeq
where $c = n_s + n_f$, with $n_s$ and $n_f$ being the numbers of free scalar and fermion fields respectively. 
It is derived from the counterterm
\beq\label{Counterterm2D}
\Gamma_{\textrm{Ct}}[g] = - \frac{\mu^{\epsilon}}{\epsilon}\, \frac{c}{24\,\pi} \,\int d^d x\, \sqrt{g}\, R\, , 
\quad \epsilon = d - 2\, .
\eeq
The Weyl gauging procedure  for the integral of the scalar curvature gives
\beq \label{GaugeCT2D}
-\frac{\mu^{\epsilon}}{\epsilon}\, \int d^dx \, \sqrt{\hat g}\, \hat R =
- \frac{\mu^{\epsilon}}{\epsilon}\, \int d^dx \, \sqrt{g} \, R + 
\int d^2 x \, \sqrt{g}\, \left[ \frac{\tau}{\Lambda}\, R + \frac{1}{\Lambda^2}\, \left(\pd\tau\right)^2 \right]\, .
\eeq
The second term in (\ref{GaugeCT2D}) is, modulo a constant, the Wess-Zumino action in $2$ dimensions,
\beq
\Gamma_{WZ}[g,\tau] = - \frac{c}{24\,\pi}\, \int d^2x\, \sqrt{g}\, \left[
\frac{\tau}{\Lambda}\, R  + \frac{1}{\Lambda^2}\, \left(\pd\tau\right)^2 \right] \, ,
\eeq
from which we can extract the 2-dilaton amplitude according to (\ref{FuncDiffWZ})
\beq
\mathcal I_{2}(\ku\,-\ku) = \frac{1}{\Lambda^2}\, \langle  T(\ku) T(-\ku) \rangle = \frac{c}{12\,\pi}\, \ku^2 \, .
\eeq
Starting from the 2-dilaton vertex, which is the only non-vanishing one, 
exploiting (\ref{NoInt}) and inverting the remaining relations, we get the Green functions
\bea
\langle T(\ku) T(\kd) T(\kt) \rangle 
&=& 
- \frac{c}{6\,\pi}\, \left( \ku^2 + \kd^2 + \kt^2 \right)\, ,
\nn \\
\langle T(\ku) T(\kd) T(\kt) T(\kq) \rangle 
&=&
\frac{c}{\pi}\, \left( \ku^2 + \kd^2 + \kt^2 + \kq^2 \right)\, , \nn \\
\langle T(\ku) T(\kd) T(\kt) T(\kq) T(\kc) \rangle 
&=&
-\frac{8\,c}{\pi}\, \left( \ku^2 + \kd^2 + \kt^2 + \kq^2 + \kc^2 \right)\, , \nn \\
\langle T(\ku) T(\kd) T(\kt) T(\kq) T(\kc) T(\ks) \rangle 
&=&
\frac{80\,c}{\pi}\, \left( \ku^2 + \kd^2 + \kt^2 + \kq^2 + \kc^2 + \ks^2 \right)\, .
\eea
These results exactly agree with the combinations of completely traced multiple functional derivatives of the anomaly
(\ref{TraceAnomaly2D}) that one derives from (\ref{hier}), providing a consistency check of our recursive formulas
(\ref{DilIntStructure6}).

\subsection{The first six traced correlators in d=4}
Here we just report the expressions of the first 6 correlators in $d=4$
Given the anomaly equation in 4 dimensions (\ref{TraceAnomaly4d}), we obtain %
\bea
\langle T(\ku) T(-\ku) \rangle 
&=& 
- 4\, \beta_a\, {\ku}^4 \, , 
\nn \\
\langle T(\ku) T(\kd) T(\kt) \rangle
&=& 
8 \bigg[ 
- \bigg( \beta_a+\beta_b \bigg)\,\bigg( \sum_{i=1}^3 k_i^2\, \left(\sum \right)
f_{3}(\ku,\kd,\kt)+f_{3}(\kd,\ku,\kt)+f_{3}(\kt,\ku,\kd)\bigg) \nn \\
&&
+\, \beta_a\, \sum_{i=1}^{3} k_i^4 \bigg]\, ,
\nn
\eea
\bea
\langle T(\ku) T(\kd) T(\kt) T(\kq)\rangle 
&=&
8\, \bigg\{ 
6\, \bigg( \beta_a + \beta_b \bigg)\, \bigg[
\sum_{\mathcal T\left\{4,[(k_{i_1},k_{i_2}),(k_{i_3},k_{i_4})]\right\}} k_{i_i}\cdot k_{i_2}\, k_{i_3}\cdot k_{i_4}
\nn \\
&&
+\, f_{4}(\ku\,\kd,\kt,\kq) + f_{4}(\kd\,\ku,\kt,\kq) + f_{4}(\kt\,\ku,\kd,\kq) + f_{4}(\kq\,\ku,\kd,\kt) \bigg]
\nn \\
&&
-\, \beta_a\, 
\bigg( \sum_{\mathcal T\left\{4,(k_{i_1},k_{i_2})\right\}}(k_{i_1} + k_{i_2})^4 + 4\, \sum_{i=1}^{4} k_{i}^4 \bigg)
\bigg\}\, , 
\nn
\eea
\bea
&&
\langle T(\ku) T(\kd) T(\kt) T(\kq) T(\kc) \rangle 
=
16\, \Biggl\{
-24\,\bigg(\beta_a + \beta_b\bigg)\, \Biggl[
\sum_{\mathcal T\left\{5,[(k_{i_1},k_{i_2}),(k_{i_3},k_{i_4})]\right\}} k_{i_1}\cdot k_{i_2}\,k_{i_3}\cdot k_{i_4}
\nn \\
&& \hspace{-10mm}
+\, f_{5}(\ku,\kd,\kt,\kq,\kc) + f_{5}(\kd,\ku,\kt,\kq,\kc) + f_{5}(\kt,\ku,\kd,\kq,\kc) 
+   f_{5}(\kq,\ku,\kd,\kt,\kc) + f_{5}(\kc,\ku,\kd,\kt,\kq) \Biggr]
\nn \\
&& \hspace{-10mm}
+\, \beta_a\, \Biggl[
                     \sum_{\mathcal T\left\{5,(k_{i_1},k_{i_2},k_{i_3})\right\}} \left(k_{i_1} + k_{i_2} + k_{i_3} \right)^4
                     + 3\, \sum_{\mathcal T\left\{5,(k_{i_1},k_{i_2})\right\}} \left(k_{i_1} + k_{i_2} \right)^4
                     + 12\, \sum_{i=1}^{5} k_{i}^4 
\Biggr]
\Biggr\}\, , \nn
\eea
\bea
&&
\langle T(\ku) T(\kd) T(\kt) T(\kq) T(\kc) T(\ks) \rangle =
32\, \Biggl\{
120\,\bigg(\beta_a + \beta_b\bigg)\, \Biggl[
\sum_{\mathcal T\left\{6,[(k_{i_1},k_{i_2}),(k_{i_3},k_{i_4})]\right\}} k_{i_1}\cdot k_{i_2}\,k_{i_3}\cdot k_{i_4} \nn \\
&& 
+\, f_{6}(\ku,\kd,\kt,\kq,\kc,\ks) + f_{6}(\kd,\ku,\kt,\kq,\kc,\ks) + f_{6}(\kt,\ku,\kd,\kq,\kc,\ks)  \nn \\
&&
+\,  f_{6}(\kq,\ku,\kd,\kt,\kc,\ks) + f_{6}(\kc,\ku,\kd,\kt,\kq,\ks) + f_{6}(\ks,\ku,\kd,\kt,\kq,\kc) \Biggr] \nn \\
&& \hspace{-10mm}
-\, \beta_a\, \Biggl[ 
\sum_{\mathcal T\left\{6,(k_{i_1},k_{i_2},k_{i_3},k_{i_4})\right\}} \left(k_{i_1} + k_{i_2} + k_{i_3}+ k_{i_4} \right)^4
+  4\, \sum_{\mathcal T\left\{6,(k_{i_1},k_{i_2},k_{i_3})\right\}} \left(k_{i_1} + k_{i_2} + k_{i_3} \right)^4
\nn \\
&&
+\,  11\, \sum_{\mathcal T\left\{6,(k_{i_1},k_{i_2})\right\}} \left(k_{i_1} + k_{i_2} \right)^4
+ 48\, \sum_{i=1}^{5} k_{i}^4 
\Biggr]
\Biggr\}\, ,
\label{BuildingBlocks}
\eea
where we have introduced the compact notation
\bea
f_{3}(k_a,k_b,k_c)
&=&
k_a^2\, k_b \cdot k_c \, ,
\nn \\
f_{4}(k_a,k_b,k_c,k_d)
&=&
k_a^2\, \left( k_b \cdot k_c + k_b \cdot k_d + k_c \cdot k_d \right)\, , 
\nn \\
f_{5}(k_a,k_b,k_c,k_d,k_e) 
&=& 
k_a^2\, \left( k_b \cdot k_c + k_b \cdot k_d + k_b \cdot k_e + k_c \cdot k_d + k_c \cdot k_e + k_d \cdot k_e  \right)\, ,
\nn \\
f_{6}(k_a,k_b,k_c,k_d,k_e,k_f)
&=& 
k_a^2\, \left( k_b \cdot k_c + k_b \cdot k_d + k_b \cdot k_e + k_b \cdot k_f + k_c \cdot k_d + k_c \cdot k_e 
\right.
\nn \\
&& \hspace{5mm}
\left.
+\, k_c \cdot k_f + k_d \cdot k_e + k_d \cdot k_f + k_e \cdot k_f  \right)\, .
\eea

\end{document}